\documentclass[fleqn,usenatbib]{mnras}

\usepackage[T1]{fontenc}
\usepackage{adjustbox}

\newcommand{\GG}[1]{}
\defcitealias{Bh+19}{Paper I}
\defcitealias{Bh+19b}{Paper II}
\defcitealias{Bh21}{Paper III}
\defcitealias{Arnaboldi22}{Paper V}

\usepackage{graphicx}	
\usepackage{adjustbox}
\usepackage{arydshln}
\usepackage{rotating}
\usepackage{longtable}
\usepackage{lscape}
\usepackage{newtxtext,newtxmath}

\title[Radial abundance gradients of the M~31 thin and thicker discs from PNe]{The survey of planetary nebulae in Andromeda (M31). IV. Radial oxygen and argon abundance gradients of the thin and thicker disc}

\author[S. Bhattacharya et al.]{
Souradeep Bhattacharya,$^{1}$\thanks{E-mail: souradeep@iucaa.in}
Magda Arnaboldi,$^{2}$ 
Nelson Caldwell,$^{3}$
Ortwin Gerhard,$^{4}$
Chiaki Kobayashi$^{5}$
\newauthor
Johanna Hartke,$^{6}$
Kenneth C. Freeman,$^{7}$
Alan W. McConnachie,$^{8}$ 
and Puragra Guhathakurta$^{9}$
\\
$^{1}$Inter University Centre for Astronomy and Astrophysics, Ganeshkhind, Post Bag 4, Pune 411007, India\\
$^{2}$European Southern Observatory, Karl-Schwarzschild-Str. 2, 85748 Garching, Germany \\ 
$^{3}$Harvard-Smithsonian Center for Astrophysics, 60 Garden Street, Cambridge, MA 02138, USA \\
$^{4}$Max-Planck-Institut für extraterrestrische Physik, Giessenbachstraße, 85748 Garching, Germany \\
$^{5}$Centre for Astrophysics Research, Department of Physics, Astronomy and Mathematics, University of Hertfordshire, Hatfield, AL10 9AB, UK\\
$^{6}$European Southern Observatory, Alonso de C\'ordova 3107, Santiago de Chile, Chile \\   
$^{7}$Research School of Astronomy and Astrophysics, Mount Stromlo Observatory, Cotter Road, ACT 2611 Weston Creek, Australia \\
$^{8}$NRC Herzberg, 5071 West Saanich Road, Victoria, BC V9E 2E7, Canada \\
$^{9}$UCO/Lick Observatory, Department of Astronomy \& Astrophysics, University of California Santa Cruz, 1156 High Street, Santa Cruz, California 95064, USA
}

\date{Accepted September 16, 2022. Received September 16, 2022; in original form February 21, 2022}

\pubyear{2022}

\begin{document}
\label{firstpage}
\pagerange{\pageref{firstpage}--\pageref{lastpage}}
\maketitle

\begin{abstract}
We obtain a magnitude-limited sample of {Andromeda (M~31)} disc PNe with chemical abundance {estimated} through the direct detection of the [\ion{O}{iii}]~4363~\AA~line. This leads to $205$ and $200$ PNe with oxygen and argon abundances respectively. We find that high- and low-extinction M~31 disc PNe have statistically distinct argon and oxygen abundance distributions. In the radial range $2-30$~kpc, the older low-extinction disc PNe are metal-poorer on average with a slightly positive radial oxygen abundance gradient ($0.006 \pm 0.003$ dex/kpc) and slightly negative {for argon} ($-0.005 \pm 0.003$ dex/kpc), while the younger high-extinction disc PNe are metal-richer on average with steeper radial abundance gradients for both oxygen ($-0.013 \pm 0.006$ dex/kpc) and argon ($-0.018 \pm 0.006$ dex/kpc), similar to the gradients {computed} for the M~31 HII regions. The M~31 {disc} abundance gradients are consistent {with values computed from major merger simulations}, with the majority of the low-extinction PNe being the older pre-merger disc stars in the thicker disc, and the majority of the high-extinction PNe being younger stars in the thin disc, formed during and after the merger event. The chemical abundance of the M~31 thicker disc has been radially homogenized because of the major merger. {Accounting for disc scale-lengths, the positive radial oxygen abundance gradient of the M~31 thicker disc is in sharp contrast to the negative one of the MW thick disc. However, the thin discs of the MW and M~31 have remarkably similar negative oxygen abundance gradients.}
\end{abstract}




\section{Introduction}
\label{sect:intro}
Late-type galaxies can contain multi-layered populations that are kinematically distinct, the ``cold" thin disc and the ``hot" thick disc, found in the Milky Way \citep[MW; e.g.][]{Gilmore83} and in nearby galaxies \citep{Yoachim06, Comeron19}. The MW thick disc {within the solar radius is found} to be chemically distinct from the MW thin disc suggesting separate evolution for the two \citep[e.g.][]{bhg16,Kobayashi20}. Thick discs may form from accreted gas during a chaotic period of hierarchical clustering at high redshift \citep{Brook04} or from dynamical heating of thinner discs by secular processes \citep{Sellwood14}. Mergers with satellites can also dynamically heat thin discs to decrease their rotational velocity and increase their velocity dispersion \citep{Quinn86}, resulting in thickened discs \citep{Hopkins09}. Even the stars from the merged satellite can form the thick disc \citep{Penarrubia06}. 

Since galaxies are thought to evolve by hierarchical mergers with satellite galaxies \citep{White78,Bullock05}, late-type galaxies of varying disc thickness are expected. {Observational evidence of multi-layered disks in nearby spirals can be obtained by measuring kinematics and metallicity of their stellar populations covering a large radial range in these discs \citep[E.g.][]{Yoachim08, Yoachim08b}. However, this requires deep spectroscopic observations of resolved stars \citep[E.g.][]{Guha05} or integral field spectroscopy \citep[E.g.][]{Saglia18}. For the nearly 12 sq. deg. on-sky coverage of the M~31 disc, such observations would be highly time consuming with current instrumentation.} 

However, spectroscopic observations for kinematics and chemistry is possible with a {reasonable investment in telescope time} with discrete tracers, such as Planetary Nebulae (PNe). PNe are discrete tracers of stellar population properties and their kinematics have been measured in galaxies of different morphological types \citep[e.g.][]{coc09, cor13, pul18, Aniyan18, hartke18, Aniyan21, Hartke22}. In particular, in \citet[][hereafter Paper II]{Bh+19b}, through velocity dispersion profiles of high- and low- extinction PNe in the M~31 disc, kinematically distinct dynamically colder thin and hotter thicker discs were respectively identified. {The thin and thicker discs of M~31 were found to have rotational velocity dispersions twice and thrice that of the MW disc population of the same age \citep[Paper II,][]{Collins11,dorman15}.} While the MW disc is thought to have evolved mainly by secular evolution \citep{Sellwood14} with its most recent merger $\sim$10 Gyr ago \citep{Belokurov18,Helmi18}, the age-velocity dispersion relation in the M~31 disc at galactocentric radial distances R$\rm_{GC}=$14--20 kpc was found to be consistent with a $\sim$1:5 major merger in M~31 $\sim$2.5--4.5 Gyr ago. {A minor merger, such as that modelled for M~31 by \citet[][]{Fardal13}, would not be able to dynamically heat the M~31 disc to the measured rotational velocity dispersion for the thicker disc \citep[E.g. see models by][]{Martig14}.} Thus, while the MW is a prime laboratory for studying effects driven by secular evolution in discs, the M~31 disc formation is considered to be primarily driven by mergers \citepalias{Bh+19b} also supported by observational evidence of substructures in its inner halo (PAndAS -- \citealt[][]{mcc09}). 

The measured rotational velocity dispersions for the M~31 thin and thicker disc is in agreement with predictions from the merger simulations by \citet{ham18}. There a gas-rich satellite was accreted on to M~31 with an orbit along the Giant Stream, heating the M~31 disc and producing a thick disc from the pre-existing stars. After the merger, the replenished cold gas would lead to the formation of a {gaseous} thin disc {with younger stars formed} through a burst of star formation. The stars in the thin and thick discs of M~31 are formed in different epochs in this scenario and so differences may be present in the chemical composition of their stars. See also the N-body simulations of a merger (mass ratio including dark matter $\sim$1:5) in M~31 by \citet{Milosevic22} as well as the suite of N-body simulations by \citet{Sadoun14}.

The chemical evolution of discs of spiral galaxies is {also} reflected in their radial abundance distributions. In an inside-out build-up scenario of a galaxy disc, negative radial metallicity gradients are expected \citep[e.g.][]{SanchezMenguiano18}. Hydrodynamical simulations have shown that the radial metallicity gradient in galaxies is modified both in case of secular evolution \citep[e.g.][]{Gibson13} and also in the case of galaxy mergers \citep[e.g.][]{Zinchenko15,Tissera19}. 

Abundance distributions in galaxies can be mapped from PNe, {as they are} tracers of chemistry in galaxies \citep{Maciel94, Kwitter21}. PN elemental abundances, particularly oxygen and argon, shed light on the ISM conditions at the time of formation of their parent stellar population \citep[e.g.][]{Bresolin10,hm11,Stanghellini14}. When PNe ages can be identified, it becomes possible to map abundance variations across different epochs of star formation in galaxies. In the MW, the radial oxygen abundance gradient for both thin and thick disc PNe formed at different epochs have been obtained separately allowing for significant constraints on its chemical evolution \citep{Stanghellini18}.

In the case of M~31, the PN oxygen abundance gradient has previously been {computed} by \citet[][]{san12} from $\sim50$ observed PNe in the M~31 disc. They found a best-fit slope of $-0.0056 \pm 0.0076$ dex/kpc within R$\rm_{GC}\sim$4--24 kpc. Later {estimates} by \citet{Kw12} (only oxygen abundances) and \citet{Pena19} (both oxygen and argon abundances) out to R$\rm_{GC}\sim110$ kpc have also found near-flat abundance gradients. The flat PN oxygen abundance gradient differs from  that {derived} for HII regions in the M~31 disc which display a much steeper best-fit slope of $-0.023 \pm 0.002$ dex/kpc \citep{Zurita12}. Since HII regions sample only a young (<0.3~Gyr) stellar population, while PNe sample a wider age range, the difference in abundance gradients of the two samples may reflect the different properties of their parent generations of stars. Elemental abundances of the high- and low- extinction PNe with distinct ages can now be {computed} to obtain separate abundance distributions of the kinematically distinct thin and thicker disc in M~31 \citepalias{Bh+19b}. 

In this paper we obtain direct {estimates} of oxygen and argon abundances for the M~31 disc PNe using a large sample size and covering a wide $2- 30$ kpc radial range. Our aim is to assess whether the kinematically distinct thin and thicker disc of M~31 have different abundance distribution and gradients. Our observations and sample selection are discussed in Section~\ref{sect:data}. The radial oxygen and argon abundance gradients of the high- and low-extinction PNe respectively, are presented in Section~\ref{sect:abund}. {We note that the radial gradient of the log(O/Ar) ratios and the distribution of PNe in the log(O/Ar) vs. 12+ log(Ar/H) planes are shown and presented in \citet[][companion paper, hereafter Paper V]{Arnaboldi22}.} We assess our radial abundance gradients in context of disc galaxies in Section~\ref{sect:disc}. We then discuss constraints on the chemical evolution and formation history of M~31 in Section~\ref{sect:disc2} and finally conclude in Section~\ref{sect:future}.

\section{Data reduction and sample selection}
\label{sect:data}

\subsection{Observations}
\label{sect:obs}

\begin{figure}
        \centering
        \includegraphics[width=\columnwidth,angle=0]{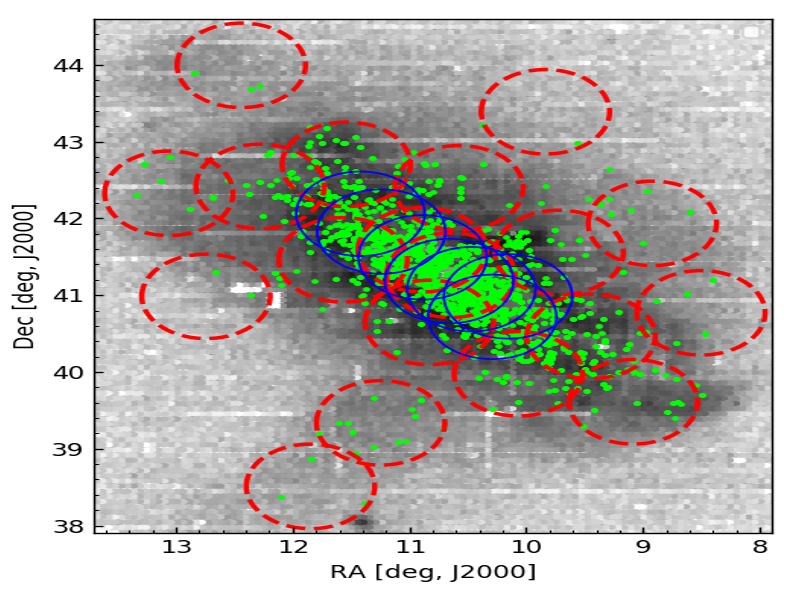}
        \caption{Position on sky of the PNe with V$\rm_{LOS}$ measurements in M31, both observed in this work as well as the archival sample from \citet{san12}, overlaid on the PAndAS number density map of RGB stars\citep{mcc18}. The targeted MMT fields are marked in red while those utilised in tag-along observations are marked in blue.}
        \label{fig:spat_pos}
\end{figure}

In \citet[][hereafter Paper I]{Bh+19}, we identified PN candidates in a 16 sq. deg. imaging survey of M~31 with MegaCam at the CFHT, covering the disc and inner halo. This was later expanded to cover 54 sq. deg in M31 \citep[][hereafter Paper III]{Bh21}.  Spectroscopic observations of a complete subsample of these PN candidates were carried out with the Hectospec multifibre positioner and spectrograph on the Multiple Mirror Telescope \citep[MMT;][]{fab05}. The Hectospec 270 gpm grating was used and provided spectral coverage from 3650 to 9200~\AA~ at a resolution of $\sim5$~\AA. Some spectra did not cover [\ion{O}{II}] 3726/3729~\AA~ because of the design of the spectrograph (alternate fibers are shifted by 30~\AA) and the small blueshift of M~31. Each Hectospec fibre subtends 1.5$''$ on the sky and was positioned on the PN candidates in each field. Table~\ref{table : obs_det} shows details of the fields observed in this work whose positions have been marked in Figure~\ref{fig:spat_pos}. We targeted 2222 distinct PNe candidates with fibres in 26 separate fields in M~31, some of which were observed multiple times. Of these fields, seven were part of tag-along observations (where only a few free fibres from other observing programs covering M31 were placed on the PN candidates). These pointings are marked with $^{*}$ in Table~\ref{table : obs_det}. 

\begin{table}
\caption{Details of MMT Hectospec observations of PNe. Brighter PNe were prioritised for observations but PNe to m$_{5007}=26.4$ mag were targeted. Some PNe were observed twice in the regions of overlap of adjacent fields. The fields for tag-along observations are marked with $^{*}$.}
\centering
\adjustbox{max width=\columnwidth}{
\begin{tabular}{c|c|c|c|c|c}
\hline
Obs. date & RA & DEC & Exp. time & N$\rm_{PN, targ}$ & N$\rm_{PN, obs}$\\
 & (deg) & (deg) & (s) &  &  \\
\hline
14.09.18$^{*}$ & 10.4700833 & 41.0778389 & 9000 & 38 & 19\\
15.09.18 & 11.2468750 & 39.3372003 & 4800 &  41 & 13\\
06.10.18$^{*}$ & 10.4700833 & 41.0778389 & 9000 & 65 & 41\\
04.12.18 & 11.5407083 & 42.7038650 & 3600 & 202 & 44\\
02.09.19 & 10.83  & 40.645 & 4800 & 148 & 67\\
05.09.19 & 10.0681250 & 39.9762458 & 3600 & 119  & 55\\
23.10.19 & 9.4446667 & 40.4633331  & 4800 & 175 & 71\\
24.10.19 & 10.8910833  & 41.5916672 & 4800 & 226 & 174\\
25.10.19 & 9.0818750 & 39.6116675 & 6000 & 79  & 26\\

07.10.20$^{*}$ & 10.659625 & 41.1844253 & 7920 & 44 & 22\\
07.10.20$^{*}$ & 10.15525 & 40.9850014 & 6600 & 53 & 21\\
08.10.20$^{*}$ & 10.8875833 & 41.4929542 & 8400 & 37& 26\\
08.10.20 & 11.568125 & 41.4599992 & 4800 & 232 & 63\\
09.10.20$^{*}$ & 10.15525 & 40.9850014 & 5280 & 88 & 34\\
09.10.20$^{*}$ & 11.2415417 & 41.8282319 & 7560 & 30 & 18\\
10.10.20$^{*}$ & 10.2915417 & 40.7175942 & 3000 & 45 & 17\\
10.10.20 & 9.7178333 & 41.5566672 & 4800 & 110 & 28\\
11.10.20$^{*}$ & 11.2415417 & 41.8282319 & 7560 & 8 & 3\\
12.10.20$^{*}$ & 11.4205 & 42.0603333 & 4500 & 26 & 14\\
12.10.20$^{*}$ & 11.4205 & 42.0603333 & 6000 & 27 & 16\\
12.10.20 & 12.2790417 & 42.4166681 & 4800 & 166 & 26\\
12.10.20 & 10.5755 & 42.3983344 & 6000 & 153 & 37\\
13.10.20 & 8.5034167 & 40.7700006 & 4800 & 26 & 8\\
13.10.20 & 12.44075 & 43.9933319 & 4800 & 13 & 5\\
24.10.20 & 10.6455667 & 41.2599983 & 3000 & 228 & 57\\
24.10.20 & 11.8477083 & 38.5099983 & 5760 & 30 & 16\\
24.10.20$^{*}$ & 10.2915417 & 40.7175942 & 4800 & 135 & 58\\
25.10.20 & 13.06125 & 42.3266678 & 6000 & 80 & 17\\
25.10.20 & 12.7429167 & 40.9850006 & 3600 & 42 & 8\\
25.10.20 & 8.919125 & 41.9350014 & 3600 & 23 & 5\\
25.10.20 & 9.8354167 & 43.3883325 & 6000 & 69 & 12\\
\hline
\end{tabular}
\label{table : obs_det}
}
\end{table}

\begin{figure*}
	\centering
	\includegraphics[width=\textwidth,angle=0]{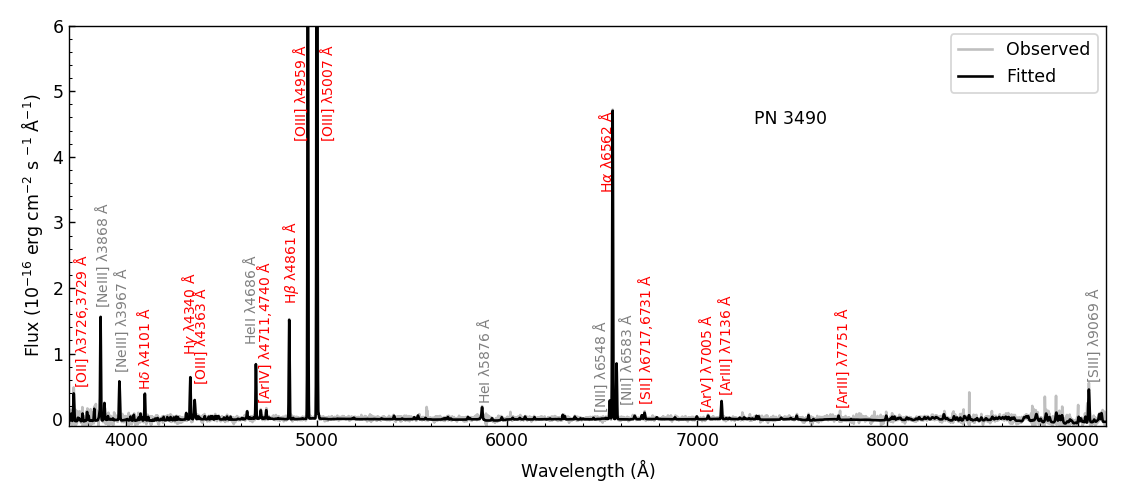}
	\caption{An example of the spectra observed by Hectospec for the PNe in M31. The spectra shown in grey is obtained following heliocentric correction, removal of sky-lines and flux calibration. The fitted spectra from ALFA \citep{Wesson16} is shown in black. The emission lines with fluxes tabulated in Table~\ref{table : linelist} are labelled in red, while other observed bright lines are marked in grey.}
	\label{fig:spec}
\end{figure*}

The initial steps for the data reduction of each Hectospec spectra are similar to that described by \citet{Caldwell09} for their observations of star clusters in M~31, which were also followed by \citet{san12} for their PN spectra. Briefly, following the de-biasing and flat-fielding of each observed field, individual spectra were extracted and wavelength calibrated, including a heliocentric correction. Standard star spectra were used for flux calibration and instrumental response. Sky subtraction was carried out by averaging spectra from fibers placed on blank sky from the same exposures or by offsetting the telescope by a few arcseconds \citep[see][]{Caldwell09}. The spectra of PN candidates that were observed multiple times (in adjacent fields) have been combined, effectively coadding those integration times. Figure~\ref{fig:spec} shows an example of the observed PN spectra. 

\subsection{Emission line fluxes and line-of-sight velocity estimation}
\label{sect:flux}

\begin{figure}
	\centering
	\includegraphics[width=\columnwidth,angle=0]{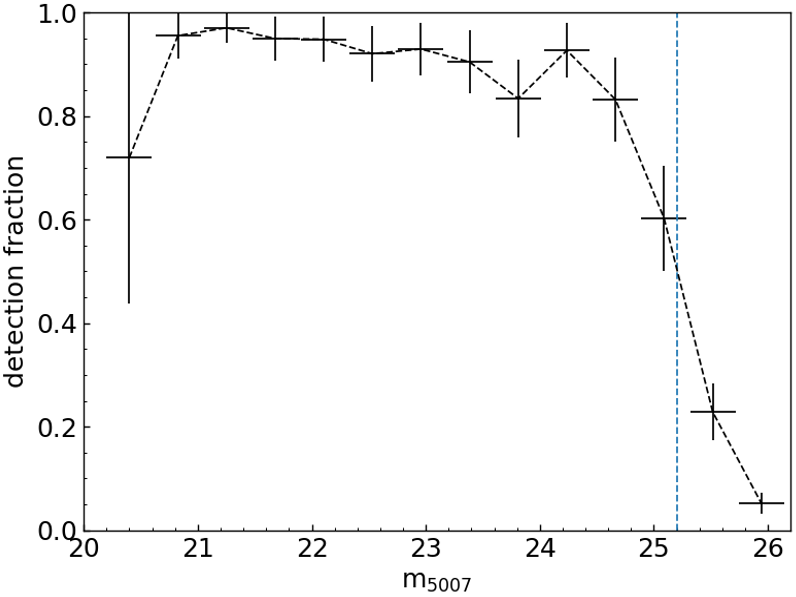}
	\caption{Fraction of PNe targeted with spectroscopic observations where [\ion{O}{iii}] 4959/5007 ~\AA~ emission lines (the [\ion{O}{iii}] doublet) {is} detected. The uncertainty in detection fraction is the binomial proportion confidence-interval of observed PNe in any magnitude bin obtained using the Wilson score interval method \citep{Wilson27}. The blue dashed line shows the 50\% detection limit of the spectroscopic follow-up.}
	\label{fig:det_frac_spec}
\end{figure}

Emission-line fluxes for each PN candidate were measured using the automated line fitting algorithm, ALFA \citep{Wesson16}, which has been tailored for emission line sources. The line-of-sight velocity (V$\rm_{LOS}$) is measured from the strongest emission-lines accounting for heliocentric correction, with an uncertainty of 3 km s$^{-1}$. After subtracting a globally-fitted continuum, ALFA derives fluxes by optimising the parameters of Gaussian fits to line profiles using a genetic algorithm. Of the 2222 targeted PNe candidates, 866 have confirmed detection of the [\ion{O}{iii}] 4959/5007 ~\AA~ emission lines. The [\ion{O}{iii}] 5007 ~\AA~ emission line is detected in all {these} cases with a signal-to-noise ratio higher than 5. All of them have H$\alpha$ emission line present also. The fraction of PNe detected as a function of magnitude is shown in Figure~\ref{fig:det_frac_spec}. We also used ALFA to obtain the emission line fluxes from the archival Hectospec spectra of 449 PNe studied by \citet{san12} which had confirmed detection of the [\ion{O}{iii}] 4959/5007 ~\AA~ emission lines. Of these archival PNe, 64 were re-observed in this work. We thus have 1251 unique PNe with V$\rm_{LOS}$ measurements in M31, termed the \textit{PN\_M31\_LOSV} sample. Their spatial position is shown in Figure~\ref{fig:spat_pos}. Note that contaminant spatially unresolved HII regions were removed and are not included in this sample (see Appendix~\ref{app:hii} for further details).

\subsection{Extinction measurement}
\label{sect:ext_mea}

For each PN, the measured emission-line fluxes are then passed to
NEAT \citep[Nebular Empirical Analysis Tool;][]{Wesson12}, which applies an empirical scheme to calculate the extinction and elemental abundances. NEAT calculated the intrinsic c(\ion{H}{$\beta$}) using the flux-weighted ratios of H$\alpha$/H$\beta$, H$\gamma$/H$\beta$ and H$\delta$/H$\beta$ (whichever pairs are observed) and the extinction law of \citet{Cardelli89}, first assuming a nebular temperature of 10000K and an electron density of 1000 cm$^{-3}$, and then recalculating c(\ion{H}{$\beta$}) at the {computed} temperature and density (whenever available; see Section~\ref{sect:abund_det} for details). Of the 1251 PNe in the \textit{PN\_M31\_LOSV} sample, 745 had the \ion{H}{$\beta$} line detected and their extinctions (A$_V$) could be determined with a positive value. Note that a further 380 PNe showed the presence of the \ion{H}{$\beta$} line but resulted in a negative (but close to zero) value of A$_V$, similar to {what was} found by \citet{san12}. These PNe were not utilised further in this work to {derive} elemental abundances but are a part of the \textit{PN\_M31\_LOSV} sample.

\begin{figure}
        \centering
        \includegraphics[width=\columnwidth,angle=0]{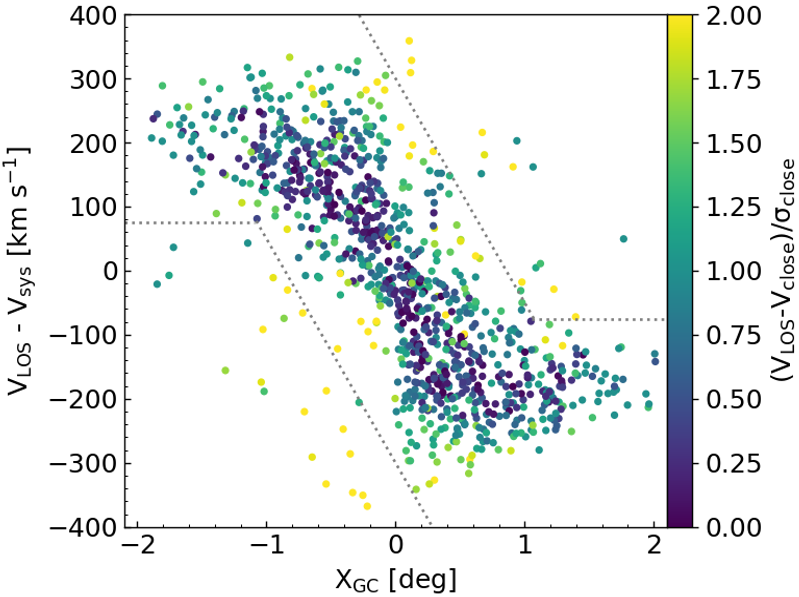}
        \caption{Position-velocity diagram of PNe within R$\rm_{GC}=30$ kpc. Here X$\rm_{GC}$ is the deprojected major-axis distance in deg (1 deg= 13.68 kpc). The dotted lines distinguish the outliers which have a non-disc angular momentum and possibly correspond to streams or halo PNe. The PNe are coloured by their (V$\rm_{LOS}$~-~V$\rm_{close}$)/$\rm \sigma_{close}$ values, where the outliers stand out in yellow. V$\rm_{close}$ and $\rm\sigma_{close}$ refer to the local LOSV and dispersion within a radius of $4'$ centred around each PN.}
        \label{fig:vel_pos}
\end{figure}

\subsection{Selection of M~31 disc PNe from the position velocity diagram}
\label{sect:pos_vel}
The PNe with extinction measurements are de-projected on to the galaxy plane based on the position angle (PA = 38$^\circ$) and inclination (i = $77^\circ$) of M~31 in the planar disc approximation. PNe beyond R$\rm_{GC}=30$ kpc are not included further in the analysis, as a significant fraction of them may be associated with the prominent bright substructures -- G1-Clump and Northern Spur, and the dwarf galaxy NGC~205, present at these radii. The remaining PNe within R$\rm_{GC}=30$ kpc are shown in Figure~\ref{fig:vel_pos} which plots their position, X$\rm_{GC}$ (de-projected major-axis distance in deg {in the disc-plane}), against their V$\rm_{LOSV}~-~V_{sys}$ (M~31 systemic velocity, V$\rm_{sys}=-309$ km s$^{-1}$; \citealt{merrett06}). While the majority of PNe in M~31 within R$\rm_{GC}=30$ kpc are associated with its bulge and disc, some PNe associated with the extension of a luminous substructure or any fainter stellar stream co-spatial with the disc may also be present. Such PNe may present themselves as a dynamically cold component that is offset from the {main disc PNe distribution} in the position-velocity plot, like the PNe associated with the extension of the Giant Stream on the disc as proposed by \citet{Merrett03}. The dotted lines in Figure~\ref{fig:vel_pos} correspond to an offset from the mean value of the V$\rm_{LOSV}~-~V_{sys}$ for the PNe to the maximum possible velocity dispersion of the thick disc in M~31 (160 km s$^{-1}$; from \citetalias{Bh+19b}). We identify the position-velocity outliers as those PNe whose V$\rm_{LOSV}~-~V_{sys}$ values are outside the maximum values for the thick disc PNe in M~31. This selection successfully identifies as outliers those PNe on the extension of the giant stream tagged previously by \citet{Merrett03} and a few additional PNe. 

A further selection of PN outliers is possible for the inner regions of the disc with large number of PN LOSV measurements, where such LOSV outliers may stand out from the LOSVs of the PNe in their local spatial neighbourhood. For each PN, we obtain the mean local LOSV, V$\rm_{close}$, and local LOSV dispersion,$\rm \sigma_{close}$, within a radius of 4$\arcmin$. The value of the radius was chosen such that there are at least 5 PNe within such radius for each PNe out to  R$\rm_{GC}\sim20$. Those PNe whose LOSV is over 2$\rm \sigma_{close}$ away from  V$\rm_{close}$, are tagged as outliers. In this way, we identify as outliers those PNe associated with the claimed extension of the giant stream from \citealt{Merrett03} as seen in Figure~\ref{fig:vel_pos}, along with additional PNe that were not classified as such in previous works. We thus identify 601 PNe with extinction measurements (see Section~\ref{sect:ext_mea}) within R$\rm_{GC}=30$ kpc which are robust  M~31 disc members (some of these within R$\rm_{GC}=5$ kpc may also belong to the M~31 bulge). This is then the M~31 disc PN sample (termed \textit{PN\_M31d\_Av} ) and Table~\ref{table : obs_npn_ext} summarises the number of PNe identified in each aforementioned step. This is the largest sample of PNe with extinction measurements observed in the M~31 disc, in fact, in any external galaxy. We have nearly doubled the sample size of PNe with extinction measurements from \citetalias{Bh+19b}. PN disc kinematics with the increased sample will be explored in a future paper. In this work, we further {focus on} the M~31 disc PN sample in order to study PN elemental abundances.


\begin{table}
\caption{Summary of numbers of PNe observed in this work and \citet{san12} to build the total sample of M~31 disc PNe with LOSV and A$\rm_{V}$ measurements.}
\centering
\adjustbox{max width=\columnwidth}{
\begin{tabular}{l|c}
\hline
\\
No. of PNe with LOSV measurement observed in this work & 866 \\
No. of \citet{san12} PNe with LOSV measurement & 449 \\
Total PNe with LOSV measurement (\textit{PN\_M31\_LOSV}) & 1251 \\
Those of the above with positive A$\rm_{V}$ measurement & 745 \\
Those of the above in the disc (\textit{PN\_M31d\_Av} ) & 601 \\
\hline
\end{tabular}
\label{table : obs_npn_ext}
}
\end{table}

\begin{figure}
	\centering
	\includegraphics[width=\columnwidth,angle=0]{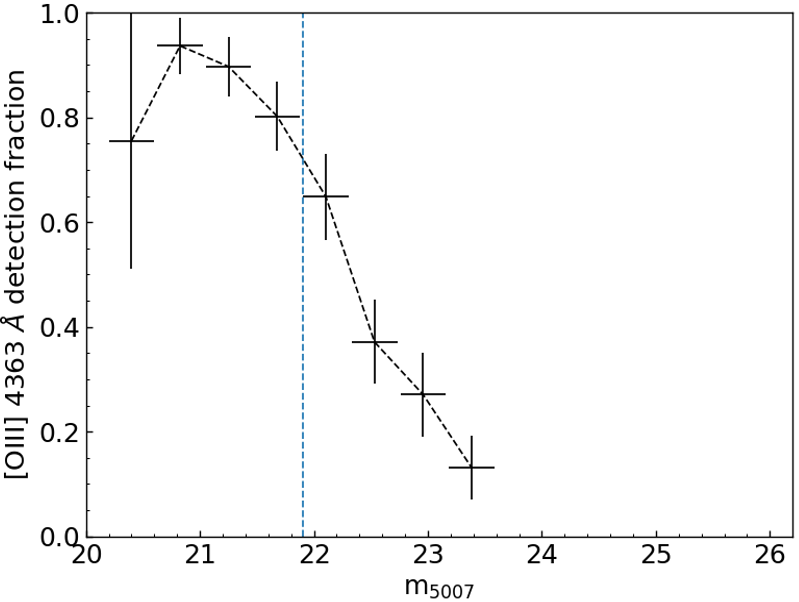}
	\caption{Fraction of PNe with spectroscopic observations where the [\ion{O}{iii}] 4363 ~\AA~ emission line {is} detected and the elemental abundances could be {derived}. The uncertainty in detection fraction is the binomial proportion confidence-interval of observed PNe in any magnitude bin obtained using the Wilson score interval method \citep{Wilson27}. The blue dashed line shows the 75\% detection limit of the [\ion{O}{iii}] 4363 ~\AA~ emission line.}
	\label{fig:det_frac_oxy}
\end{figure}

\begin{figure}
        \centering
        \includegraphics[width=\columnwidth,angle=0]{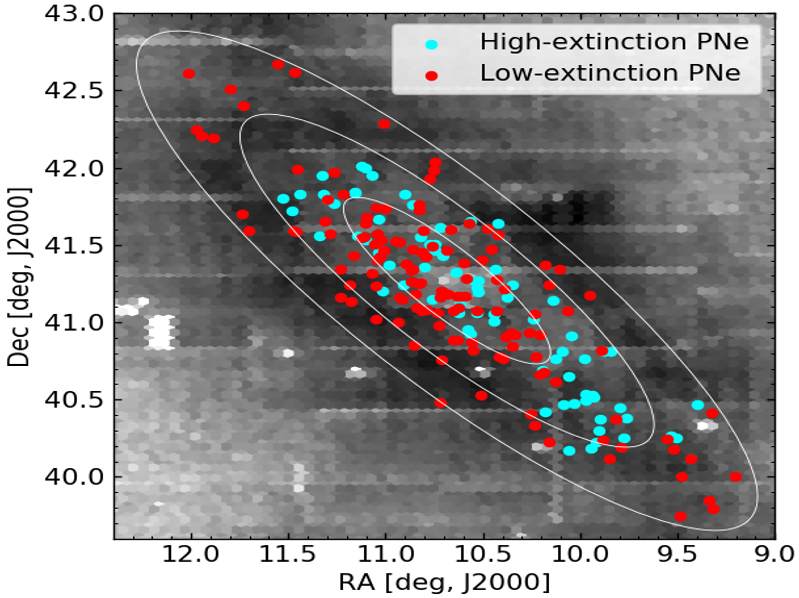}
        \caption{Position on sky of the PNe with oxygen abundances {determined} in M31, overlaid on the PAndAS number density map of RGB stars\citep{mcc18}. The high- and low-extinction PNe are marked in cyan and red respectively. The white ellipses show R$\rm_{GC}=10,20,30$~kpc respectively.}
        \label{fig:spat_oxy}
\end{figure}

\begin{table*}
\caption{Properties of the 205 M~31 PNe in the \textit{PN\_M31d\_O\_lim} sample. Column 1: Serial number of the PN in this work. Following IAU naming conventions, each PN should be designated as SPNA$<$Sl. No.$>$. E.g. PN 419 should be termed SPNA419; Column 2--3: spatial position of the PN; Column 4: LOSV of the PN; Column 5: Measured balmer decrement of the PN and corresponding extinction in Column 6; Column 7--8: {Derived} abundances of the PN; Column 9: The [\ion{O}{III}] 5007 ~\AA~ magnitude measured in \citetalias{Bh+19} and \citetalias{Bh21}. A portion of this table is shown here for guidance; the full table will be made available through the CDS.}
\centering
\adjustbox{max width=\textwidth}{
\begin{tabular}{cccccccccccc}
\hline
Sl. No. & RA [J2000] & DEC [J2000] & V$\rm_{LOS}$ & c(\ion{H}{$\beta$}) & A$\rm_{V}$ & T$\rm_{e}$ & N$\rm_{e}$  & 12+log(O/H) & 12+log(Ar/H) & m$_{5007}$\\
 & deg & deg & km s$^{-1}$ & mag & mag & K & cm$^{-3}$  &  &  & mag \\
\hline
478 & 11.3247088 & 41.9470492 & -223.5 & 0.68 & 1.44 & $12100^{+400}_{-400}$ & $418^{+945}_{-417}$ & 8.52 $\pm$ 0.04 & 6.28 $\pm$ 0.07 & 21.2\\
496 & 11.2640144 & 41.9713806 & -125.7 & 0.32 & 0.69 & $11800^{+400}_{-400}$ & $2450^{+955}_{-1360}$ & 8.68 $\pm$ 0.1 & 6.39 $\pm$ 0.05 &  21.03\\
942 & 12.0147552 & 42.610521 & -102.0 & 0.18 & 0.37 & $11300^{+500}_{-500}$ & $1000$ & 8.35 $\pm$ 0.05 & 6.02 $\pm$ 0.04 &  21.2\\
945 & 11.4699967 & 42.6147482 & -174.0 & 0.09 & 0.18 & $10600^{+300}_{-300}$ & $2040^{+1030}_{-590}$ & 8.52 $\pm$ 0.04 & 6.15 $\pm$ 0.03 &  20.82\\
959 & 11.5577759 & 42.6747226 & -83.8 & 0.24 & 0.51 & $10100^{+300}_{-300}$ & $2560^{+1450}_{-800}$ & 8.7 $\pm$ 0.2 & 6.16 $\pm$ 0.04 &  21.02\\
\hline
\end{tabular}
\label{table : prop}
}
\end{table*}

\subsection{Direct determination of elemental abundances for each PN}
\label{sect:abund_det}

Emission lines in the spectra of each PN of the \textit{PN\_M31d\_Av}  sample are de-reddened using the calculated c(\ion{H}{$\beta$}) and their temperatures and densities are calculated using an iterative process from the relevant diagnostic lines using NEAT \citep[see][section 3.3]{Wesson12}. For our observations, NEAT utilizes the temperature-sensitive [\ion{O}{iii}] 4363 \AA~line\footnote{{Note that the temperature sensitive [\ion{N}{II}]~5755~\AA~line is also observed for some PNe in our sample. However, the derived [\ion{N}{II}]-based temperature is more uncertain than the [\ion{O}{III}]-based temperature while also being inherently less reliable (see Appendix~\ref{app:temp}). Thus only the [\ion{O}{III}]-based temperature is used for chemical abundance computation in this work.}} and the density-sensitive [\ion{O}{ii}] 3726/3729 \AA~ and [\ion{S}{ii}] 6717/6731 \AA~ doublets to obtain temperature and electron density for each PN, whenever the [\ion{O}{iii}] 4363 \AA~ line\footnote{{We ignore recombination contribution to the [\ion{O}{iii}] 4363 \AA~ line as the relevant recombination lines are much too faint to be observed at the distance of M~31 with present day instrumentation/telescopes.}} is observed. Oxygen and argon ionic abundances are {derived} from the observed fluxes of the oxygen ([\ion{O}{II}] 3726/3729 \AA,  [\ion{O}{III}] 4363/4959/5007~\AA) and argon ([\ion{Ar}{III}] 7136/7751 \AA, [\ion{Ar}{IV}] 4711/4740 \AA, [\ion{Ar}{V}] 7005~\AA) {emission} lines respectively. Elemental oxygen and argon abundances are obtained from the ionic abundances using the ionisation correction factors (ICFs) from \citet{Delgado14} {where applicable following the default NEAT prescription (see Appendix~\ref{app:icf} for details).} Uncertainties are propagated through all steps of the analysis into the final values. Comparison with archival PN abundance determinations is discussed in Appendix~\ref{app:lit}. Of the 601 PNe in the \textit{PN\_M31d\_Av}  sample, 276 have oxygen abundances {determined}, out of which 269 also have argon abundances. 

Figure~\ref{fig:det_frac_oxy} shows the detection fraction of the temperature-sensitive [\ion{O}{iii}] 4363 ~\AA~ emission line, enabling the {computation} of oxygen and argon abundances, {for} those PNe with spectroscopic observations. The [\ion{O}{iii}] 4363 ~\AA~ emission line, relative to the [\ion{O}{iii}] doublet, is brighter for lower metallicity PNe. Hence, we are more likely {to detect the [\ion{O}{iii}]~4363~\AA~emission line} in metal-poor PNe at the faint end. We thus restrict the analysis of the abundance distribution and gradient to those PNe where the detection fraction is higher than 75\%, i.e, m$_{5007}\leq21.9$mag, so as not to be biased towards metal-poor PNe while still maintaining a large sample, 205 and 200 PNe with oxygen and argon abundance {values determined} respectively. The magnitude-limited sample of 205 M~31 disc PNe with oxygen abundances is termed the \textit{PN\_M31d\_O\_lim} sample. It is the sample for which the analysis is carried out in this work. Their spatial position is shown in Figure~\ref{fig:spat_oxy}. The emission line fluxes of the lines of interest for these PNe have been listed, along with their 1$\rm\sigma$ uncertainties, in Table~\ref{table : linelist}. Their measured V$\rm_{LOS}$, A$_V$, oxygen and argon abundances are listed in Table~\ref{table : prop} as well as their observed m$_{5007}$ magnitudes obtained in \citetalias{Bh+19} and \citetalias{Bh21}. In Appendix~\ref{app:lit} we compare our independently {determined} oxygen abundance values with those in the literature for previously observed individual M31 PN spectra. 
Table~\ref{table : obs_npn} summarises the selection of the magnitude-limited abundance sample of M~31 disc PNe from the \textit{PN\_M31d\_Av} sample. This is the largest sample of PNe with chemical abundances {derived} in M~31, and in fact, in any external galaxy. 

\begin{table}
\caption{Summary of PN selection to build the \textit{PN\_M31d\_O\_lim} sample used in this work from the \textit{PN\_M31d\_Av} sample. We also summarise the extinction classification of the \textit{PN\_M31d\_O\_lim} sample discussed in Section~\ref{sect:ext}.}
\centering
\adjustbox{max width=\columnwidth}{
\begin{tabular}{l|c}
\hline
\\
M~31 disc PN with oxygen abundance  & 276 \\
M~31 disc PN with argon abundance  & 269 \\
Magnitude-limited sample with oxygen abundance (\textit{PN\_M31d\_O\_lim}) & 205 \\
Those of the previous with argon abundance & 200 \\
\hdashline
\\
High-extinction PNe in the \textit{PN\_M31d\_O\_lim} & 75 \\
Those of the previous with argon abundance & 75 \\
\hdashline
\\
Low-extinction PNe in the \textit{PN\_M31d\_O\_lim} & 130 \\
Those of the previous with argon abundance & 125 \\
\hline
\end{tabular}
\label{table : obs_npn}
}
\end{table}


\subsection{Classification of planetary nebulae based on extinction measurements}
\label{sect:ext}

\begin{figure}
        \centering
        \includegraphics[width=\columnwidth,angle=0]{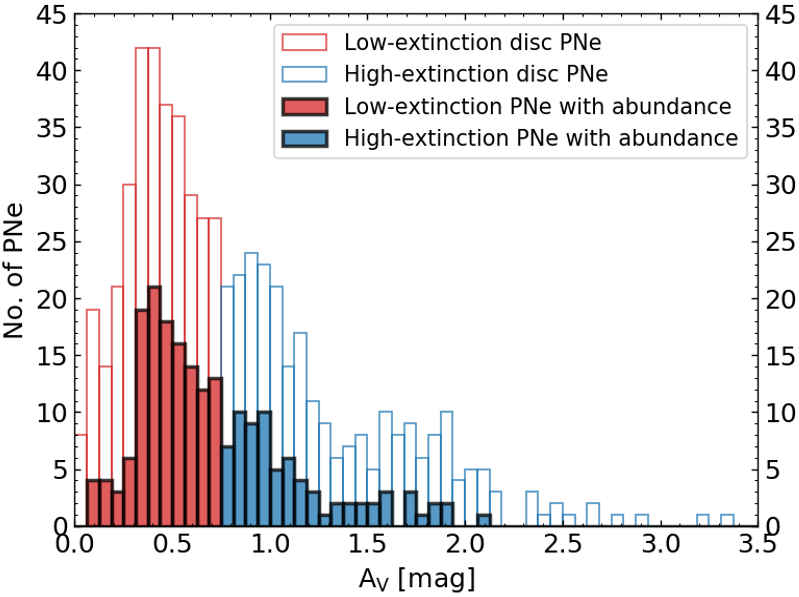}
        \caption{Histogram (shaded) showing the distribution of extinction values of the PNe in the \textit{PN\_M31d\_O\_lim} sample (magnitude limited sample with oxygen abundances {determined}). The high- and low-extinction PNe lie in the blue and red shaded regions respectively. The histogram (unshaded) shows the distribution of extinction values of the \textit{PN\_M31d\_Av} sample for comparison.}
        \label{fig:av_hist}
\end{figure}

The mass of PN central stars correlates with their circumstellar extinction \citep{Ciardullo99}. This is because dust production of stars in the AGB phase scales exponentially with their initial progenitor masses for the $1\sim2.5 M_{\odot}$ range after which it remains roughly constant \citep{Ventura14}. Additionally, PNe with dusty high-mass progenitors evolve faster \citep{MillerB16} and so their circumstellar matter has little time to disperse, while PNe with lower central star masses evolve sufficiently slowly that a larger fraction of dust is dissipated from their envelopes \citep{Ciardullo99}. In \citetalias{Bh+19b}, the high- and low-extinction PNe constituted the kinematically distinct thin and thicker disc of M~31 respectively. From archival CLOUDY photoionization models \citep{cloudy} of a subsample of these PNe \citep{Kw12}, we found ages of $\sim2.5$  and $\sim4.5$ Gyr for the high- and low-extinction PNe  respectively. As in \citetalias{Bh+19b}, based on the distribution of the M31 PNe extinction values which exhibits a drop at A$\rm_V = 0.75$ mag, we classify PNe with extinction values higher and lower than A$\rm_V = 0.75$ mag as high- and low-extinction PNe respectively \citepalias[for further details, see Section 3.1 in][]{Bh+19b}. Our \textit{PN\_M31d\_O\_lim} sample is then divided into 75 high- and 130 low-extinction PNe, which are expected to be associated with the thin and thicker disc stellar populations respectively, and with age ranges, ~ 2.5 Gyrs and younger (high-extinction PNe), and 4.5 Gyrs and older (low-extinction PNe). All the PNe have oxygen abundances and all but five low-extinction PNe also have argon abundances. Table~\ref{table : obs_npn} also summarises the number of high- and low-extinction PNe in the \textit{PN\_M31d\_O\_lim} sample. Figure~\ref{fig:av_hist} shows the distribution of extinction values of the PNe in the \textit{PN\_M31d\_O\_lim} sample, along with that for the the M~31 disc PNe sample. The high- and low-extinction PNe are marked separately in Figure~\ref{fig:spat_oxy}. As shown in Figure~\ref{fig:spat_oxy} and further discussed in Section~\ref{sect:abund}, the high-extinction PNe are concentrated within a smaller radial extent than the low-extinction PNe.


\begin{figure}
        \centering
        \includegraphics[width=\columnwidth,angle=0]{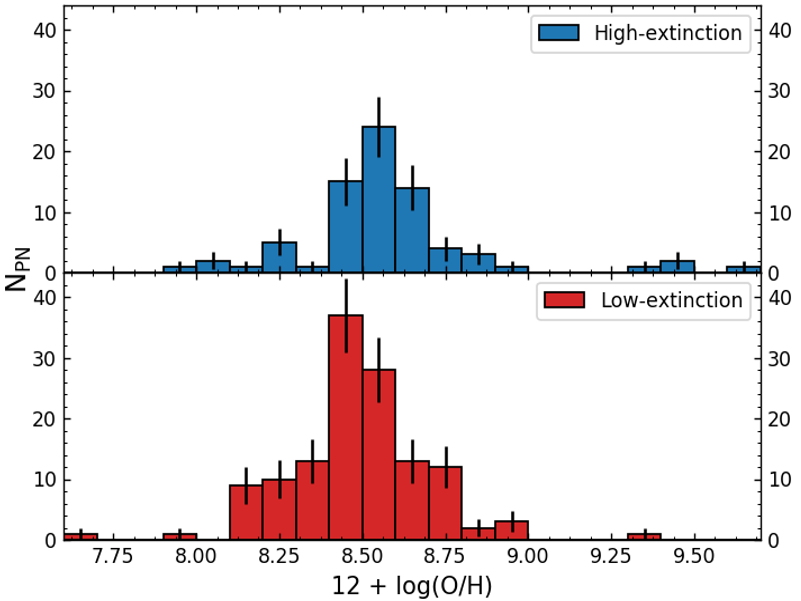}
        \caption{Histogram showing the distribution of oxygen abundances for the [top] high- and [bottom] low-extinction PNe in the \textit{PN\_M31d\_O\_lim} sample. The bins are 0.1 dex wide, {vertical bars indicate} the Poissonian errors.}
        \label{fig:met_hist}
\end{figure}

\section{Abundance distribution and radial gradients in the M~31 disc from Planetary Nebulae}
\label{sect:abund}

We separately explore the chemical abundance distribution and gradients of both oxygen and argon for the high- and low-extinction PNe which trace the kinematically distinct thin and thicker discs of M~31. We note that \citet{Delgado-Inglada15} suggested a possible dependency of PN oxygen abundance on metallicity and mass for carbon dust rich (CRD) MW PNe. As per their study, CRD MW PNe may be oxygen enriched by up to 0.3 dex from modification of surface oxygen in the Asymptotic Giant Branch (AGB) phase. However, argon has been found to be invariant during the AGB phase, providing an independent probe to the ISM conditions at the time of birth of the PN parent stellar population. We, however, do not find that oxygen is enriched in CRDs in a larger MW PNe sample from \citet{Ventura17}, detailed in Appendix~\ref{sect:agb}. Additionally, for PNe evolving from stars with initial mass $\geq3\rm M_{\odot}$ (younger than $\sim300$~Myr), hot-bottom burning (HBB) may result in an oxygen depletion of up to $\sim0.2$ dex. The M~31 high- extinction PNe have average ages $\sim2.5$~Gyr with the bulk of them having likely formed in a burst of star formation $\sim$2 Gyr ago (\citetalias{Bh+19b}). This implies that a very small number of PNe with very young massive progenitors (affected by HBB) are expected in our sample\footnote{{In \citetalias{Arnaboldi22}, we studied the log(O/Ar) versus 12+log(Ar/H) PN distribution as function of log(N/Ar). We conclude that the log(O/Ar) values did not show any dependency on log(N/Ar) with the implication that the fraction of massive ($\geq3\rm M_{\odot}$) young ($\sim300$~Myr and younger) PNe in the M31 sample is very small.}}. The bulk of the M~31 PNe would thus exhibit oxygen and argon abundances unaffected by AGB evolution. Hence we proceed to use oxygen and argon abundances for high- and low-extinction PNe in M31 as probes of the chemical abundances of the ISM at the time of their birth for the different kinematic components of the M~31 disc. 


\subsection{Oxygen abundance distribution and radial gradient from Planetary Nebulae}
\label{sect:oxy}

Figure~\ref{fig:met_hist} shows the PN oxygen abundance distribution for all the high- and low-extinction PNe in the \textit{PN\_M31d\_O\_lim} sample of the M~31 disc. The mean value of the oxygen abundance for all the high-extinction PNe, $\rm <12+(O/H)>_{high-ext}=8.57 \pm 0.03$, is higher than that for all the low- extinction PNe, $\rm <12+(O/H)>_{low-ext}=8.48 \pm 0.02$. While the high-extinction PN sample has higher oxygen abundance on average than the low-extinction sample, there is a considerable overlap in the distributions which is reflected in the large standard deviation values 
$\rm \sigma(12+(O/H))_{high-ext}=0.28$ and $\rm \sigma(12+(O/H))_{low-ext}=0.21$. We can establish whether the oxygen abundance distributions of the two PN samples are different by statistically comparing them. We utilize the two-sample Anderson-Darling test \citep[AD-test;][]{ksample_ADtest} to compare the two distributions, which yields a significance level of 2.3\%. Since the significance level is lower than 5\%, the null hypothesis that the two samples are drawn from the same distribution is rejected. The difference in the oxygen abundance distributions of the two PN populations stems from the distinct ISM conditions at the time of their birth.

\begin{figure}
        \centering
        \includegraphics[width=\columnwidth,angle=0]{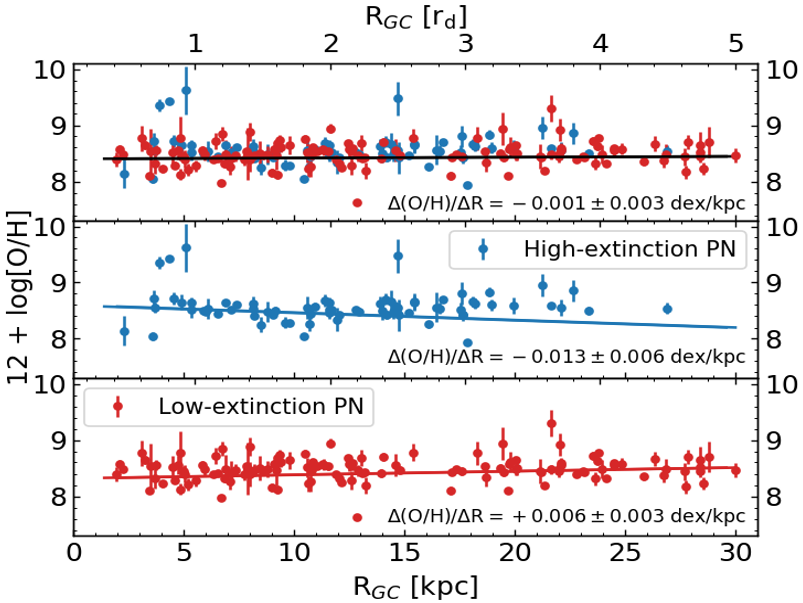}
        \caption{The galactocentric radial distribution of oxygen abundances for [top] all, [middle] high- and [bottom] low-extinction PNe in M~31 in our \textit{PN\_M31d\_O\_lim} sample. {The bottom and top axes show R$\rm_{GC}$ in units of kpc and r$\rm_{d}$ respectively.} {High- and low-extinction PNe are shown in blue and red respectively.} The best-fitting radial oxygen abundance gradient to the \textit{PN\_M31d\_O\_lim} sample is shown for all ({black}), high- (blue) and low-extinction (red) PNe. }
        \label{fig:met_grad}
\end{figure}

Figure~\ref{fig:met_grad} shows the galactocentric radial distribution of PN oxygen abundances for all PNe, as well as distinctly for high- and low-extinction PNe, in the \textit{PN\_M31d\_O\_lim} sample. We note that the high-extinction PNe cover a smaller radial extent than the low-extinction PNe, i.e. the thicker disc in M~31 is more radially extended than the thin disc {(see also \citealt{Yoachim06} for more examples of thin discs embedded in thicker discs)}. The best-fit parameters are noted in Table~\ref{table : oxyfit}. For the \textit{PN\_M31d\_O\_lim} sample, we find a near-flat abundance gradient for all PNe, $\rm(\Delta (O/H)/\Delta R)_{all} = 0.001 \pm 0.003$ dex/kpc, a steeply negative oxygen abundance gradient for the high-extinction PNe, $\rm(\Delta (O/H)/\Delta R)_{high-ext} = -0.013 \pm 0.006$ dex/kpc while for the low-extinction PNe we find a {slightly} positive oxygen abundance gradient, $\rm(\Delta (O/H)/\Delta R)_{low-ext} = 0.006 \pm 0.003$ dex/kpc. \citet{Zurita12} found that for HII regions the oxygen abundance gradient (determined indirectly for a large sample), $\rm(\Delta (O/H)/\Delta R)_{HII-regions} =-0.023 \pm 0.002$ dex/kpc is steeper than that observed for either of the PN populations. However, we do find a similar intercept for the high-extinction PNe (12+log(O/H)$\rm_{0, ~high-ext}=8.6 \pm 0.08$) and the HII regions (12+log(O/H)$\rm_{0, ~HII-regions}=8.72 \pm 0.18$; \citealt{Zurita12}). Both high extinction PNe and HII regions show the highest oxygen abundance values at R$\rm_{GC} < 5$ kpc.

\begin{table}
\caption{Fitted parameters for radial gradients in the M~31 disc from the \textit{PN\_M31d\_O\_lim} sample.}
\centering
\adjustbox{max width=\columnwidth}{
\begin{tabular}{ccccc}
\hline
PN sample & X & X$\rm_0$ & \multicolumn{2}{c}{$\rm\Delta X/\Delta R$}\\
 &  &  & dex/kpc & dex/r$\rm_{d}$\\
\hline
All & 12+log(O/H) & 8.4 $\pm$ 0.04 & 0.001 $\pm$ 0.003 & 0.006 $\pm$ 0.018 \\
High-extinction & 12+log(O/H) & 8.6 $\pm$ 0.08 & -0.013 $\pm$ 0.006 & -0.079 $\pm$ 0.036  \\
Low-extinction & 12+log(O/H) & 8.31 $\pm$ 0.05 & 0.006 $\pm$ 0.003 & 0.036 $\pm$ 0.018  \\
\\
\hdashline
All & 12+log(Ar/H) & 6.37 $\pm$ 0.04 & -0.008 $\pm$ 0.002 & -0.049 $\pm$ 0.012  \\
High-extinction & 12+log(Ar/H) & 6.51 $\pm$ 0.08 & -0.018 $\pm$ 0.006 & -0.109 $\pm$ 0.036  \\
Low-extinction & 12+log(Ar/H) & 6.3 $\pm$ 0.05 & -0.005 $\pm$ 0.003 & -0.03 $\pm$ 0.018  \\
\hline
\end{tabular}
\label{table : oxyfit}
}
\end{table}

\begin{figure}
        \centering
        \includegraphics[width=\columnwidth,angle=0]{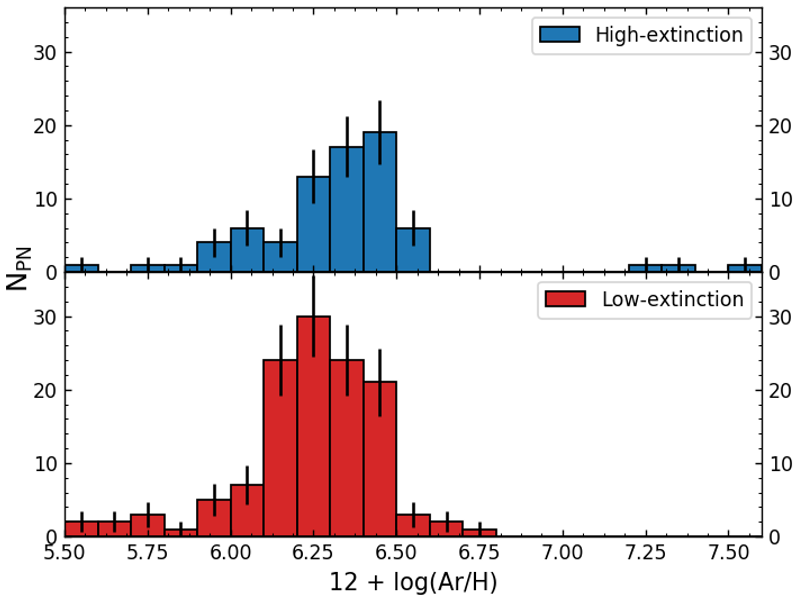}
        \caption{Histogram showing the distribution of argon abundances for the [top] high- and [bottom] low-extinction PNe. The bins are 0.1 dex wide, vertical bars indicate the Poissonian errors.}
        \label{fig:arg_hist}
\end{figure}

\begin{figure}
        \centering
        \includegraphics[width=\columnwidth,angle=0]{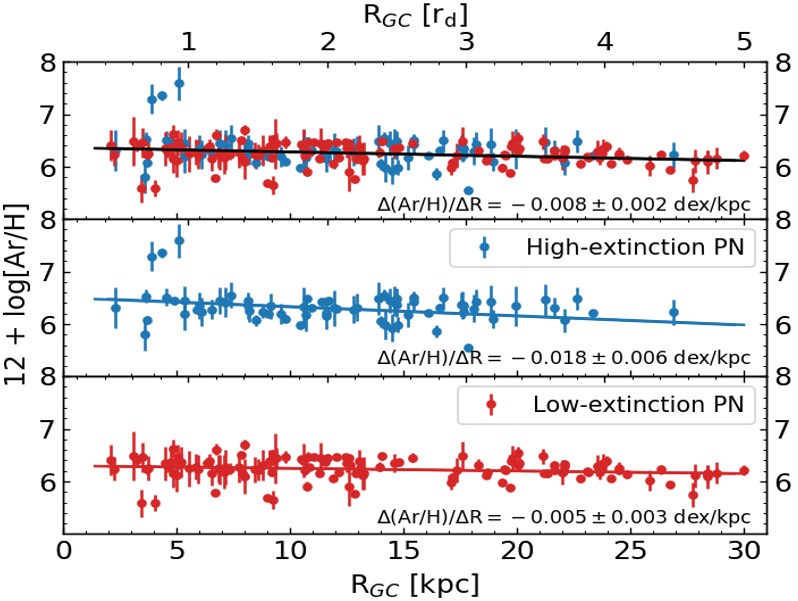}
        \caption{The galactocentric radial distribution of argon abundances for [top] all, [middle] high- and [bottom] low-extinction PNe in the M~31 disc in our \textit{PN\_M31d\_O\_lim} sample. {The bottom and top axes show R$\rm_{GC}$ in units of kpc and r$\rm_{d}$ respectively.} {High- and low-extinction PNe are shown in blue and red respectively.} The best-fitting radial argon abundance gradient to the \textit{PN\_M31d\_O\_lim} sample is shown for all ({black}), high- (blue) and low-extinction (red) PNe. }
        \label{fig:arg_grad}
\end{figure}

\subsection{Argon abundance distribution and radial gradient from Planetary Nebulae}
\label{sect:arg}

Figure~\ref{fig:arg_hist} shows the PN argon abundance distribution for the high- and low-extinction PNe in the M~31 disc within $\sim$30 kpc. We utilize the AD-test to statistically compare the two distributions which yields a significance level of 3.3\%. Since the significance level is lesser than 5\%, the null hypothesis that the two samples are drawn from the same distribution is rejected. Thus the two disc components in M31 traced by the high- and low- extinction PNe have stellar populations that were born from ISM with distinct argon and oxygen abundance distributions. Their parent stellar populations, forming the thin and thicker disc of M~31, are not only kinematically distinct \citepalias[][]{Bh+19b}, but also have distinct elemental abundances and radial gradients. The mean value of the argon abundance for the high-extinction PNe, $\rm <12+log(Ar/H)>_{high-ext}=6.32 \pm 0.03$, is clearly higher than that for the low- extinction PNe, $\rm <12+log(Ar/H)>_{low-ext}=6.25 \pm 0.02$. The standard deviation values, $\rm \sigma(12+log(Ar/H))_{high-ext}=0.29$ and $\rm \sigma(12+log(Ar/H))_{low-ext}=0.2$, reflect an overlap of their argon abundance distribution. 

Figure~\ref{fig:arg_grad} shows the galactocentric radial distribution of PN argon abundances for the high- and low-extinction PNe samples in R$\rm_{GC}=$2--30 kpc radial range. Their fitted parameters are also noted in Table~\ref{table : oxyfit}. We find a steeply negative argon abundance gradient for the high-extinction PNe, $\rm(\Delta (Ar/H)/\Delta R)_{high-ext} = -0.018 \pm 0.006$ dex/kpc while for all PNe and the low-extinction PNe we find {negative and slightly negative argon abundance gradients respectively}, $\rm(\Delta (Ar/H)/\Delta R)_{all} = -0.008 \pm 0.002$ dex/kpc and $\rm(\Delta (Ar/H)/\Delta R)_{low-ext} = -0.005 \pm 0.003$ dex/kpc. 

We can then compare the argon abundance gradients in PNe with that of the oxygen abundance gradient {computed} for HII regions\footnote{{Direct determination of oxygen and argon abundances for HII regions has been carried out only for 16 HII regions (where the faint [\ion{O}{iii}] 4363 ~\AA~ emission line is observed; \citealt{Esteban20}) over a limited radial range. This small sample and limited radial range do not allow for a well constrained determination of radial oxygen and argon abundance gradient for HII regions in M~31. However, as mentioned in Section~\ref{sect:oxy}, a much larger sample of HII regions have oxygen abundances determined using strong line methods \citep{Zurita12}, giving a more robust estimate of the oxygen abundance gradient for HII regions in M~31, suitable for comparison with the gradients determined from PNe.}}. The high-extinction PNe have an abundance gradient which is consistent within error with that of the HII regions, $\rm(\Delta (O/H)/\Delta R)_{HII-regions} =-0.023 \pm 0.002$ dex/kpc \citep{Zurita12}, implying that both the parent stellar populations of these young PNe and the HII regions originated from a similarly-enriched ISM. The near-flat abundance gradient of the low-extinction PNe, on the other hand, implies a parent stellar population which has a different chemical composition and radial trend.


\section{Comparison of radial abundance gradient from M~31 disc Planetary Nebulae with gradients
in galaxy discs}
\label{sect:disc}

\subsection{Comparison with previous nebular abundance gradient {estimates} in the M~31 disc}
\label{sect:comp_PNe_M31}

Previous PN abundance gradient {estimates} in the M~31 disc \citep{san12,Kw12,Pena19} {computed} a flat radial oxygen abundance gradient out to large galactocentric radii. \citet{Pena19} found an oxygen abundance gradient of $-0.001 \pm 0.001$ dex/kpc within R$\rm_{GC} \sim$110 kpc, including both disc and halo PNe and an argon abundance gradient of $-0.002 \pm 0.001$ dex/kpc for the same sample. The flat oxygen abundance gradient determined in this work for all PNe ($0.001 \pm 0.003$ dex/kpc; Table~\ref{table : oxyfit}) is consistent with what was previously {computed}. Our derived radial argon abundance gradient for all PNe of $-0.008 \pm 0.002$ dex/kpc within R$\rm_{GC} \sim$30 kpc is steeper than that observed by \citet{Pena19}. Given their larger galactocentric radial range and smaller PNe sample size, coupled with fewer high-extinction PNe at large radii the comparison of their argon radial abundance gradient is more suitable with our derived argon radial abundance gradient for low-extinction PNe ($-0.005 \pm 0.003$ dex/kpc). Both these values (ours and \citealt{Pena19}) are consistent within errors. 

{The} previous PN abundance gradient {estimates} in the M~31 disc were in contrast to the steeper gradient computed from HII regions \citep{Zurita12}. By classifying PNe samples on the basis of their intrinsic extinction, we have identified that the younger, dynamically colder \citepalias{Bh+19b}, high-extinction PNe have a {steeper} radial oxygen abundance gradient {which is} consistent with that of the HII regions. We confirm that the older dynamically hotter low-extinction PNe have {a flat oxygen abundance} gradient that also drives the abundance gradient of all PNe jointly given their larger number. 

\subsection{Comparison with previous stellar metallicity gradients in the M~31 disc}
\label{sect:comp_star_M31}

\begin{figure}
        \centering
        \includegraphics[width=\columnwidth,angle=0]{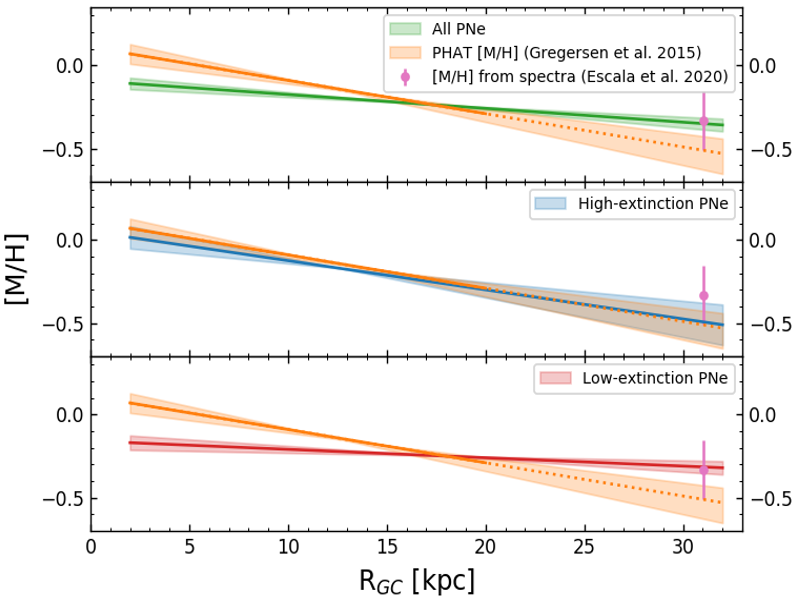}
        \caption{The best-fitting radial argon abundance gradient to the \textit{PN\_M31d\_O\_lim} sample, scaled to [M/H], is shown for [top] all, [middle] high- and [bottom] low-extinction PNe. In each panel, we show the best-fit radial gradient to the PHAT photometric metallicity (fitted to R$_{GC}=20$ kpc by \citealt{Gregersen15}, shown as a solid orange line, and extrapolated beyond as dotted line). The uncertainty in the fits are shaded. Also shown in each panel is the spectroscopic [M/H] obtained by \citet{Escala20} for resolved RGB stars in a small field.}
        \label{fig:stellar}
\end{figure}

We can compare the radial oxygen and argon abundance gradient to other stellar metallicity {gradients derived} in the M~31 disc. We decide to convert {the} argon abundance gradient to stellar metallicity  as argon is unequivocally invariant during AGB evolution (see Appendix~\ref{sect:agb} for details). To facilitate comparison, the argon abundance is converted to [M/H] by subtracting the solar [Ar/H] value (=6.38; \citealt{Asplund21}) as is done for calibrating stellar and gas phase mass–metallicity relations of galaxies (e.g., \citealt{Zahid17}). This has no effect on the gradients. We now compare the metallicity gradient from PN argon abundances with other chemical information {derived} from stars in the M~31 disc from independent studies. \citet{Gregersen15} found a metallicity gradient of $-0.02 \pm 0.004$ dex/kpc from R$\rm_{GC}\sim$4--20 kpc in the M~31 disc in the PHAT survey. They assumed solar [$\alpha$/Fe] and a constant red giant branch (RGB) age of 4~Gyr. Note that choosing a different age had consequences on the [M/H] intercept but the metallicity gradient remained unchanged within errors. Figure~\ref{fig:stellar} shows the metallicity gradient found by \citet{Gregersen15} in each panel. Their fitted stellar metallicity gradient is most similar to that of the high-extinction PNe, which are estimated to have a younger age of $\sim2.5$ Gyr \citepalias{Bh+19b}. This implies that their RGB stars assumed to have a mean age of 4~Gyr, may be contaminated by younger stars. We further note that \citet{Saglia18}, from IFU observations of the M~31 central regions within R$\rm_{GC}\sim$5 kpc, found a [M/H] gradient of $0.0 \pm 0.03$~dex/kpc. Given the error, their gradient is consistent with that of both the high- and low- extinction PNe. 

\citet{Escala20} {computed} [Fe/H] and [$\alpha$/Fe] values from individual stars in small field at R$\rm_{GC}\sim$31 kpc in the M~31 disc. Their mean [Fe/H] and [$\alpha$/Fe] values in this field are converted to [M/H] using the relation from \citealt{Salaris05} (see Appendix B in \citetalias{Bh21} for details) to find their [M/H]=$-0.33 \pm 0.18$. Since very few high-extinction PNe are expected to be found beyond 20 kpc radius \citepalias{Bh+19b}, it is likely that the [M/H] {derived} by \citet{Escala20} corresponds to PNe in the older thicker disc, the same population probed by the low-extinction PNe. In fact, we can clearly see in  Figure~\ref{fig:stellar} that their spectroscopic [M/H] {value} is consistent with the metallicity value obtained from the argon abundance gradient for the low extinction PNe at these radii. 

\subsection{Comparison of PN radial abundance gradients of the Milky Way and M~31}
\label{sect:comp_PNe_MW}

Considering all available MW disc PNe, \citet{Stanghellini18} {derived} radial oxygen and argon abundance gradients of $-0.021$ dex/kpc and $-0.029$ dex/kpc respectively. Adopting MW disc scale-length, r$\rm_{d}=2.3$~kpc from \citet{Yin09}, a suitable value within the range of disc scale-length measurements \citep{bhg16}, this corresponds to oxygen and argon abundance gradients of $-0.048$ dex/r$\rm_{d}$ and $-0.067$ dex/r$\rm_{d}$ respectively. Adopting a M~31 disc scale length of r$\rm_{d}=6.08$~kpc \citep{Yin09}, the oxygen and argon abundance gradients for all PNe in the \textit{PN\_M31d\_O\_lim} sample in this work have radial gradients of $0.006\pm0.018$ dex/r$\rm_{d}$ and $-0.049\pm0.018$ dex/r$\rm_{d}$ respectively. These are noted in Table~\ref{table : oxyfit} along with those of other sub-samples calculated previously. In both the MW and M~31, PNe have flatter oxygen abundance gradients than argon ones but the oxygen abundance gradient is much flatter in M~31 than in the MW. The argon radial abundance gradient in M~31 is also consistent with that of the MW within errors. 

\begin{figure}
        \centering
        \includegraphics[width=\columnwidth,angle=0]{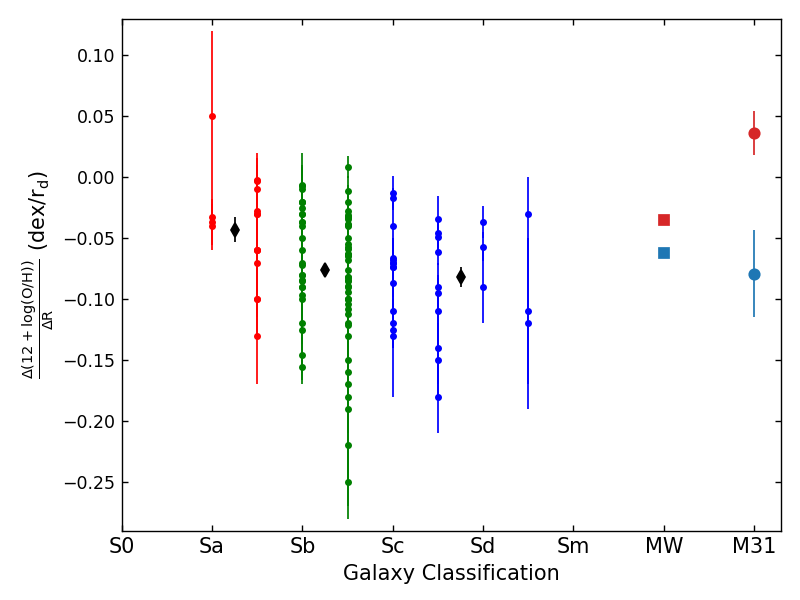}
        \caption{The oxygen radial abundance gradients {derived} in terms of disc scale length for galaxies of different morphological types in \citet{Sanchez-Menguiano16}. The Sa-Sab (red), Sb-Sbc (green) and Sc-Sdm (dark blue) galaxies also have their mean oxygen abundance gradients plotted in black. The oxygen abundance gradients for the MW \citep{Stanghellini18} (square symbols) and M~31 (this work, larger dots) are marked. The MW and M~31 thin disc values are marked in light blue while that for the MW thick disc and the M~31 thicker disc are marked in red.}
        \label{fig:morph}
\end{figure}

\citet{Stanghellini18} separately probed the thin and thick disc of the MW using younger (<1 Gyr; YPPNe) and older (>7 Gyr; OPPNe) PNe respectively. Their YPPNe exclusively populated the MW thin disc while the OPPNe had a parent stellar population dominated by thick disc stars with some contribution from old thin disc stars. They found that the YPPNe and the OPPNe had oxygen abundance gradients (in their selected sample; see their Table 4) of $-0.027$ dex/kpc and $-0.015$ dex/kpc respectively. In terms of disc scale lengths, YPPNe and the OPPNe would have oxygen abundance gradients of $-0.062$ dex/r$\rm_{d}$ and $-0.035$ dex/r$\rm_{d}$ respectively, marked in Figure~\ref{fig:morph}. These can be compared to the oxygen abundance gradients in M~31 (see Table~\ref{table : oxyfit} and the marked values in Figure~\ref{fig:morph}). The high-extinction PNe in M~31 ( with ages $\sim2.5$ Gyr; \citetalias{Bh+19b}) have a gradient in remarkable agreement with that of the YPPNe in the MW. 
The low-extinction PNe ( with ages $\sim$4.5 Gyr; \citetalias{Bh+19b}) have a positive near-flat gradient, which differs from the negative gradient of the OPPNe in the MW. The positive gradient in the M~31 thicker disc may be the end result from a major-merger event, to be discussed later in Section~\ref{sect:merg_chem}. 

\subsection{Comparison with radial abundance gradients of other galaxies}
\label{sect:comp_other}

\citet{Sanchez-Menguiano16} {computed} the oxygen abundance profiles in a sample of 122 face-on spiral galaxies observed by the CALIFA IFU survey\citep{Sanchez12} using both binned spaxels and individually identified HII regions\footnote{{We note that an updated study of galaxy properties from an expanded CALIFA IFU survey has been presented by \citet{EspinosaPonce22}. However, they do not provide the oxygen abundance gradient of individual galaxies in their sample but only their mean properties as a function of mass and morphological classification, which are in broad agreement with the results of \citet{Sanchez-Menguiano16}. As seen in Figure~\ref{fig:morph}, the diversity of galaxy radial oxygen abundance gradients is not entirely captured by their mean values. For a meaningful comparison to the oxygen abundance gradient of the MW and M~31, individual galaxy oxygen abundance gradients are required with any binned sample needing to take the properties of the MW and M~31 into account. We therefore limit our comparison to the smaller CALIFA sample of \citet{Sanchez-Menguiano16}.}}. Figure~\ref{fig:morph} shows the oxygen abundance gradients in terms of disc scale length for galaxies of different morphological types in their sample. Sa--Sab type galaxies that have the least prominent spiral arms with lowest star-formation, also show the flattest radial oxygen gradients. Other morphological types of spiral galaxies (Sb--Sdm) in the CALIFA sample have relatively steeper abundance gradients with a mean of $-0.07$ dex/r$\rm_{d}$. Such individual spirals span a wide range of radial oxygen gradients values, with many having near-flat gradients like those of the MW thick disc, while several others having steeper slopes than that observed for the MW and M~31 thin disc. However, the thicker disc of M~31 has a positive radial oxygen gradient comparable only to one {such value} for a Sa-type galaxy in the CALIFA sample, being flatter than any of those for the Sab--Sdm type galaxies.

\section{Constraints on the formation history of the M~31 disc}
\label{sect:disc2}

\subsection{Inferences on chemical evolution of galaxies from radial abundance gradients}
\label{sect:merg_chem}

Simulations of chemical evolution in isolated galaxies make predictions on the variation of the radial abundance gradient over time depending on the choice of physical mechanisms, particularly feedback prescriptions, that govern the enrichment of elements into the ISM \citep[e.g.][]{Gibson13,Molla19}. Such simulations generally predict either an initial flat gradient that steepens over time or an initial steep one that flattens over time \citep{Gibson13, Molla19}. One can attempt to constrain such models by comparing with radial abundance gradients of stars formed at different epochs in a galaxy, as carried out using the PNe, HII regions and other stellar tracers in the MW by \citet{Stanghellini18} and \citet{Molla19}. 

To constrain the chemical evolution of galaxies from chemical abundance gradients, {estimates} are required for at least two epochs. The gradients for the high and low extinction PN samples provide {estimates} in M~31 for two distinct but relatively broad age ranges, $\sim$2.5~Gyr and $\sim$4.5 Gyr and older respectively. A third epoch for such a comparisons, the present-day epoch, is provided by the oxygen abundance gradient in the M~31 HII regions. The thicker disc PNe (corresponding to a redshift, z$\sim$0.5) have a flatter abundance gradient than the thin disc PNe or the HII regions. The chemical evolution models of isolated galaxies in \citet{Molla19} predict a gradient of $-0.0106\pm 0.0010$~dex/kpc (or $-0.053 \pm 0.005$~dex/r$\rm_{d}$) for the MW at z$\sim$0.5 (\citealt{Gibson13} predict $-0.04$~dex/r$\rm_{d}$ for the same), much steeper than the observed positive radial gradient ($0.036 \pm 0.018$~dex/r$\rm_{d}$) of the M~31 thicker disc. In fact, such a positive radial gradient is not predicted in any models of chemical evolution for isolated disc galaxies covering a wide range of total masses \citep{Molla05}, which have their flattest radial oxygen gradient at $-0.01$~dex/r$\rm_{d}$. \footnote{The radial oxygen gradient of the M~31 thin disc ($-0.079 \pm 0.036$~dex/r$\rm_{d}$) is comparable with that of chemical evolution models of MW-type isolated galaxies ($-0.1$~dex/r$\rm_{d}$; \citealt{Molla19}).}

\subsection{Radial migration as a driver for the flat oxygen gradient in the M31 thicker disc}
\label{sect:radmigr}

\citet{Magrini16} noted that the radial oxygen abundance gradient of the M~31 PNe derived from the entire sample of \citet{san12} was flatter than the predictions from the chemical evolution models of isolated galaxies by \citet{Molla05}. Since such models do not account for the dynamical effects of secular evolution, \citet{Magrini16} attributed the {flat radial} oxygen abundance gradient in the M~31 disc to radial migration. Radial migration, possibly induced by bar resonances and transient spiral arms, may displace stars from their birth positions to larger radii thereby flattening the radial abundance gradient \citep{Roskar08,Minchev11}. 

Is then radial migration a possible explanation for the positive abundance gradient of the M31 thicker disc? If we were to focus only on the redistribution of stars at larger radii, then indeed radial migration brings about a flattening of the abundance gradient. According to \citet{Sellwood14} such secular processes do not dynamically heat the disc though, and thus one can not reproduce the observed high rotational velocity dispersion ($\rm\sigma_{\phi}= 101\pm 13$ km s$^{-1}$; \citetalias{Bh+19b}) of the thicker disc PNe, which would then be left unexplained.

Since merger events do flatten the radial metallicity gradients of pre-merger discs \citep{Zinchenko15} and can heat the discs \citep{Quinn86}, the {derived} flatter (even positive for oxygen) abundance gradients of the M~31 thicker and dynamically hot disc, compared to that from chemical evolution models of isolated discs, {shows} the influence of the recent merger on the radial metallicity gradient of M~31. It is further explored in the next section. 

\subsection{The radial elemental abundance gradient in galaxy merger simulations and the merger scenario in M~31}
\label{sect:merg_inf}

N-body simulations of interacting galaxies have shown that mergers leave imprints on the metallicity gradient of a galaxy, including dilution of the concentration of metals in the central part of galaxies due to gas inflow during initial passages as well as flattening of the radial metallicity gradient during the interaction \citep[][]{rupke10,Zinchenko15}. 
A near-flat abundance gradient has also been seen in EAGLE cosmological simulations of disc galaxies which experienced mergers with mass ratio $\geq$1:10 \citep{Tissera19}.  

In a minor merger scenario in M~31 as advocated by \citet{Fardal13}, a satellite galaxy (mass ratio $\sim$ 1:20) infalls along the giant stream on to the M~31 disc $\sim$1 Gyr ago. Such a satellite however would not be able to produce a heated disc with the velocity dispersion of  $~100$ km s$^{-1}$ as measured in \citetalias{Bh+19b} for the low-extinction PNe in M~31 and would additionally not form a distinct hot thin disc \citep{Martig14}. Following the major merger scenario described by \citet{ham18}, however, the pre-merger disc in M~31 would be perturbed by the a massive satellite (mass ratio $>$ 1:4.5) in a highly retrograde orbit. A prediction of this merger model is that a thin disk is rebuilt from the gas brought in by the satellite along with a burst of star formation following the dissolution of {said} satellite.

\citet{Zinchenko15} quantified the effect of mergers on the radial elemental abundance profiles of MW mass galaxies using N-body simulations (no new star formation). They found that the amount of flattening of the radial abundance gradient at large radii depends on the mass and inclination of the in-falling satellite, with flatter gradients observed for the more massive mergers. They find the maximum possible flattening from $\sim$1:20 and $\sim$1:6 mergers are 0.041~dex/r$\rm_{d}$ and $0.067$~dex/r$\rm_{d}$ respectively which occur for prograde mergers. {We can check the flattening of the radial oxygen abundance gradient if} we assume that the pre-merger thicker disc of M~31 had a radial oxygen abundance gradient similar to that of the MW ($-0.035$ dex/r$\rm_{d}$), a reasonable assumption given the thin discs of the two galaxies have similar radial gradients (see Section~\ref{sect:comp_PNe_MW}). {Then the M~31 thicker disc gradient was flattened by $0.071$~dex/r$\rm_{d}$, consistent with a mass ratio} of the merger event in M~31 of at least $\sim$1:6 or larger depending on the orbital inclination of the infalling satellite.

The observed high-extinction PNe are $\sim$2.5 Gyr old \citepalias[or younger;][]{Bh+19b} and likely trace the thin disc during its formation. The chemical evolution of an isolated thin disc after its formation has been shown in other hydrodynamic models \citep[e.g.][]{Molla19} to result in a negative radial abundance gradient, consistent with the observed negative abundance gradient for the thin disc high-extinction PNe. Furthermore, \citet{Vincenzo20} show that {a starburst following} gas in-fall into a galaxy at large radii can steepen the metallicity gradient of stars formed after the in-fall. In the case of M~31, the steep thin disc radial oxygen and argon abundance gradients are consistent with the thin disc having formed {in a starburst event} from less enriched gas brought in by the satellite mixed with the enriched ISM in the pre-merger M31 disc within R$_{GC}=14$~kpc\footnote{For details see the galactic chemical evolution models for the inner M~31 disc, presented in \citetalias{Arnaboldi22}, with the loop in the log(O/Ar) vs. 12+log(A/H) plane for the young higher extinction PNe at this radii.} \citepalias{Arnaboldi22}. The stars in the thin disc at R$_{GC}>18$~kpc are formed {in a starburst event} predominantly from the satellite gas \citepalias{Arnaboldi22}. The dilution of the ISM in the M~31 disc from the accreted satellite gas is consistent with the low stellar metallicity \citep{Conn16,Cohen18} of the giant stream substructure, which is the remaining stellar trail left by the satellite. In this context, see the recent results on the giant stream metallicity distribution from N-body simulations by \citet{Milosevic22}.

The elemental abundance gradient from PNe thus acts as constraints for merger-induced chemical evolution simulations in galaxies in general and M~31 in particular. While the fairly major merger simulations by \citet{ham18} do predict the formation of distinct thin and thick discs as observed, predictions of the abundance gradients from such simulations (not explicitly predicted in \citealt{ham18}) must be constrained in future investigations using the current values {derived} for the ISM using PNe.


\section{Conclusions}
\label{sect:future}
We present the largest sample of PNe in the M~31 disc with extinction measurements, oxygen abundances and argon abundances. We classify our observed PNe on the basis of their measured extinction. Oxygen and argon abundance distributions and radial gradients are derived for the high- and low-extinction PNe separately in the M~31 disc from direct temperature measurements. The high- and low- extinction PN abundance gradients trace the younger thin and older thicker disc of M~31 respectively. Our conclusions can be summarised as follows:
\begin{itemize}
    \item Comparing the oxygen and argon abundances in the thin and thicker discs of M~31 reveals that the two discs have distinct abundance distributions. This is the first evidence of chemically distinct thin and thicker discs in M~31.
    \item We find a steeper radial abundance gradient for the thin disc of M~31 (consistent with that of HII regions) and a near-flat (slightly positive for oxygen and slightly negative for argon) abundance gradient for the thicker disc. This is also consistent with the findings of previous studies whose near-flat PN abundance gradient estimates were dominated by the more numerous low-extinction PNe.
    \item The steep abundance gradient of the M~31 thin disc is consistent with the younger thin disc having been formed following a wet merger event. The chemical enrichment history of the thin and thicker disc of M~31 has been explored in \citetalias{Arnaboldi22} through the log(O/Ar) vs 12+log(Ar/H) plane, using the oxygen and argon abundances determined in this work. There we found the M~31 thicker disc had an extended star-formation history while the thin disc formed in a burst of star-formation {following} a wet merger event with metal-poor gas brought in by the satellite. 
    \item The thin discs of the MW and M~31 have remarkably similar oxygen abundance gradients when difference in their disc-scale lengths is taken into account.
    \item The abundance gradients for the thicker disc is flatter than expected from chemical evolution models of isolated galaxies but are consistent with the expectations of a major merger scenario (mass ratio $\sim$1:5; \citetalias{Bh+19b}). The oxygen abundance gradient of the M~31 thicker disc, in particular, is amongst the most positive observed till date in spiral galaxies, much more positive than that of the MW thick disc. The chemical abundance of the M~31 thicker disc has been radially homogenised as a consequence of the merger event. Given that the merger mass and orbital inclination has measurable influence on the metallicity gradient \citep{Zinchenko15}, the observed abundance gradients can provide constraints on the mass and inclination of the merging satellite in a major-merger scenario in M~31.

\end{itemize}

\section*{Acknowledgements}
We thank the anonymous referee for their comments. SB acknowledges support from the European Southern Observatory (ESO), Garching, Germany during his PhD. A preliminary version of this work appears in his PhD thesis \citep{PhDthesis}. SB is funded by the INSPIRE Faculty award (DST/INSPIRE/04/2020/002224), Department of Science and Technology (DST), Government of India. MAR and SB thank ESO for supporting SB’s visit through the 2021 ESO SSDF. MAR, SB and OG are grateful for the hospitality of the Mount Stromlo Observatory and the Australian National University (ANU). MAR and OG thank the Research School of Astronomy and Astrophysics at ANU for support through their Distinguished Visitor Program. This work was supported by the DAAD under the Australia-Germany joint research program with funds from the German Federal Ministry for Education and Research. CK acknowledges funding from the UK Science and Technology Facility Council through grants ST/R000905/1 and ST/V000632/1. Based on observations obtained at the MMT Observatory, a joint facility of the Smithsonian Institution and the University of Arizona. Based on observations obtained with MegaPrime/MegaCam, a joint project of CFHT and CEA/DAPNIA, at the Canada-France-Hawaii Telescope (CFHT). This research made use of Astropy-- a community-developed core Python package for Astronomy \citep{Rob13}, SciPy \citep{scipy}, NumPy \citep{numpy} and Matplotlib \citep{matplotlib}. This research also made use of NASA’s Astrophysics Data System (ADS\footnote{\url{https://ui.adsabs.harvard.edu}}).

\section*{Data Availability}
Tables~\ref{table : prop} and \ref{table : linelist} provide the required data on the kinematics and chemical abundances of the M~31 \textit{PN\_M31d\_O\_lim} PN sample and will be made available in full through the CDS. The PN spectra can be shared upon reasonable request to the authors.



\bibliographystyle{mnras}
\bibliography{ref_pne} 



\appendix

\section{HII region contamination}
\label{app:hii}

\begin{figure}
        \centering
        \includegraphics[width=\columnwidth,angle=0]{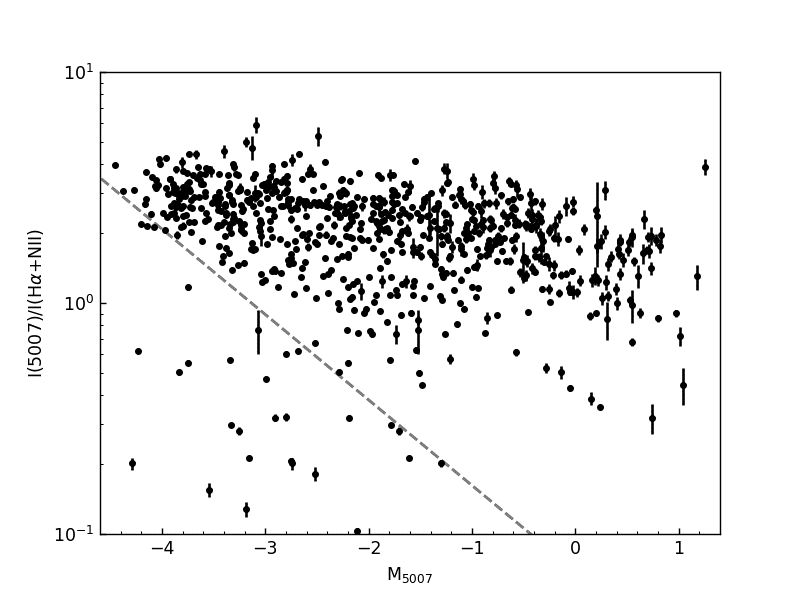}
        \caption{The ratio of foreground extinction corrected fluxes of  [\ion{O}{iii}] 5007 ~\AA~ and H-$\alpha$+NII for the PNe candidates, where the NII flux is assumed to be its maximum value of half the H-$\alpha$ flux. The dashed line separates HII regions from PNe.}
        \label{fig:hii}
\end{figure}

The PNe are identified in \citetalias{Bh+19} (and later in \citetalias{Bh21}) as point-like sources bright in the [\ion{O}{iii}] narrow-band but faint in the broad g-band. With the high angular resolution of Megacam coupled with good seeing conditions, most HII regions at the distance of M~31 appear extended, as verified from HST observations \citep[PHAT][]{dal12} in the M~31 disc in \citetalias{Bh+19}. However, contamination from ultra-compact HII regions (of radius < 10 pc) is still possible. From the PN spectra, we can distinguish between PNe and HII regions following the method described by \citet{ciardullo02} and \citet{Herrmann08} using the flux ratio of the [\ion{O}{iii}] 5007 ~\AA~ and H-$\alpha$ line (and NII which can be a maximum of 50\% of the H-$\alpha$ line; \citealt{Kreckel17}) as a function of absolute magnitude. For the candidate PNe identified in this work and including the archival sample from \citet{san12}, we plot in Figure~\ref{fig:hii} the aforementioned line ratio, corrected for foreground extinction (A$\rm_V=0.19$ mag; \citealt{Schlegel98}), against the absolute narrow-band magnitude, M$_{5007}$ observed in \citetalias{Bh+19} for the distance \citep[773 kpc;][]{Conn16} and foreground extinction of M31. The dashed line shows the selection criteria for PNe described by \citet{ciardullo02} and \citet{Herrmann08}, the region below which is occupied by HII regions. Thirty of the [\ion{O}{iii}] emitting sources may be classified as HII regions by this criteria. The remaining 1251 candidates are bonafide PNe.  

\section{Flux measurements and abundance estimates}
\label{app:measure}

\subsection{Measured line fluxes}
\label{app:flux}
The measured line fluxes for the PNe in the \textit{PN\_M31d\_O\_lim} sample are noted in Table~\ref{table : linelist}.

\begin{table*}
\caption{Measured line fluxes of the 205 M31 PNe in the \textit{PN\_M31d\_O\_lim} sample. Column 1 shows the Sl. No. of the PN in this work while the latter columns refer to the observed flux of different emission lines, relative to \ion{H}{$\beta$}=100. Following IAU naming conventions, each PN should be designated as SPNA$<$Sl. No.$>$. E.g. PN 478 should be termed SPNA478. A portion of this table is shown here for guidance; the full table will be made available through the CDS.}
\centering
\adjustbox{max width=\textwidth}{
\begin{tabular}{cccccccccccccccc}
\hline
Sl. No. & [\ion{O}{II}] & [\ion{O}{II}] & \ion{H}{$\delta$} & \ion{H}{$\gamma$} & [\ion{O}{III}] & [\ion{Ar}{IV}] & [\ion{Ar}{IV}]  & [\ion{O}{III}] & [\ion{O}{III}] & \ion{H}{$\alpha$}  & [\ion{S}{II}] & [\ion{S}{II}] & [\ion{Ar}{V}] & [\ion{Ar}{III}] & [\ion{Ar}{III}]\\
 & 3726 \AA & 3729 \AA & 4102 \AA & 4340 \AA & 4363 \AA & 4711 \AA & 4740 \AA & 4959 \AA & 5007 \AA & 6562 \AA  & 6717 \AA & 6731 \AA & 7005 \AA & 7136 \AA & 7751 \AA \\
\hline
\\
478 & 17.7 $\pm$ 4.9 & 18.8 $\pm$ 4.8 & 19.3 $\pm$ 2.5 & 17.4 $\pm$ 1.6 & 15.0 $\pm$ 1.2 & -- & 5.1 $\pm$ 1.3 & 581.9 $\pm$ 6.6 & 1753.9 $\pm$ 18.0 & 428.0 $\pm$ 6.2 & -- & -- & -- & 28.0 $\pm$ 1.7 & 16.3 $\pm$ 16.3 \\
496 & 36.7 $\pm$ 4.5 & 36.1 $\pm$ 4.6 & 22.7 $\pm$ 2.3 & 27.1 $\pm$ 1.9 & 17.3 $\pm$ 1.5 & 5.9 $\pm$ 1.4 & 8.1 $\pm$ 1.2 & 600.7 $\pm$ 7.8 & 1792.4 $\pm$ 21.7 & 331.1 $\pm$ 10.2 & 5.1 $\pm$ 1.1 & 9.2 $\pm$ 1.0 & -- & 25.2 $\pm$ 1.4 & 15.1 $\pm$ 15.1 \\
942 & -- & -- & 28.1 $\pm$ 1.1 & 25.5 $\pm$ 0.8 & 8.2 $\pm$ 1.0 & -- & -- & 309.6 $\pm$ 3.9 & 963.3 $\pm$ 13.6 & 296.3 $\pm$ 3.7 & -- & -- & -- & 10.5 $\pm$ 0.2 & 5.4 $\pm$ 5.4 \\
945 & 32.1 $\pm$ 2.4 & 21.8 $\pm$ 2.4 & -- & 34.8 $\pm$ 1.1 & 8.3 $\pm$ 0.6 & -- & 1.5 $\pm$ 0.3 & 366.9 $\pm$ 3.5 & 1105.7 $\pm$ 13.7 & 277.5 $\pm$ 4.4 & 2.1 $\pm$ 0.2 & 3.2 $\pm$ 0.3 & -- & 12.8 $\pm$ 0.4 & 6.9 $\pm$ 6.9 \\
959 & 34.0 $\pm$ 2.0 & 19.8 $\pm$ 2.3 & -- & 20.2 $\pm$ 0.8 & 8.5 $\pm$ 0.6 & -- & -- & 455.7 $\pm$ 5.1 & 1339.6 $\pm$ 13.9 & 311.3 $\pm$ 5.8 & 1.9 $\pm$ 0.3 & 3.9 $\pm$ 0.4 & -- & 14.5 $\pm$ 0.4 & 7.3 $\pm$ 7.3 \\
\\
\hline
\end{tabular}
\label{table : linelist}
}
\end{table*}

\subsection{Comparing direct temperature from [\ion{O}{III}] and [\ion{N}{II}]}
\label{app:temp}

\begin{figure}
        \centering
        \includegraphics[width=\columnwidth,angle=0]{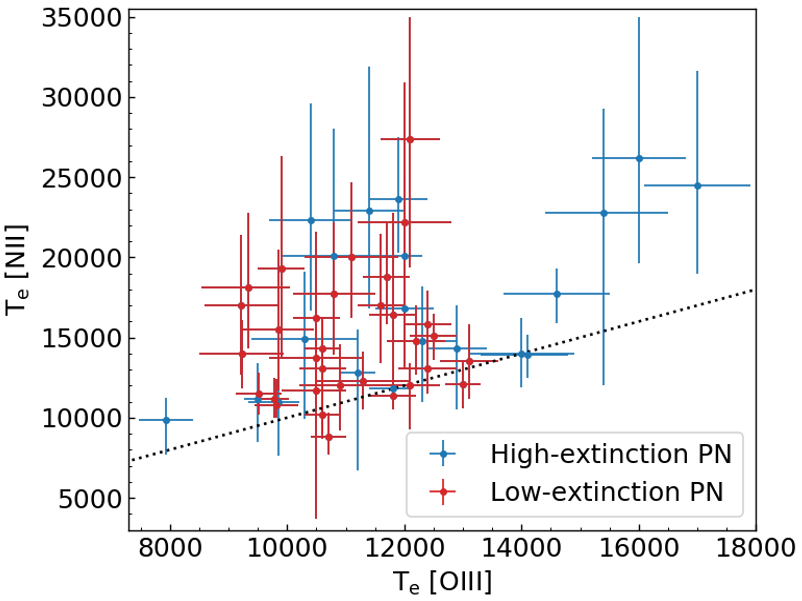}
        \caption{Comparison of the direct [\ion{O}{III}]-based temperature with the direct [\ion{N}{II}]-based temperature for those 52 PNe studied in this work where the [\ion{N}{II}]~5755~\AA~line flux has been measured. High- and low-extinction PNe are marked separately. The dotted line shows the 1:1 line.}
        \label{fig:temp}
\end{figure}

The complete sample of 205 PNe studied in this work have flux measurements of the temperature-sensitive [\ion{O}{III}]~4363~\AA~line, thereby allowing for direct [\ion{O}{III}]-based temperature (T$\rm_{e}$ [\ion{O}{III}]) determination. Of these PNe, 52 also have flux measurements of the temperature-sensitive [\ion{N}{II}]~5755~\AA~line, thereby allowing for direct [\ion{N}{II}]-based temperature (T$\rm_{e}$ [\ion{N}{II}]) determination. Figure~\ref{fig:temp} shows the comparison of the (T$\rm_{e}$ [\ion{O}{III}]) and (T$\rm_{e}$ [\ion{N}{II}]) for these 52 PNe. Note that the mean error on T$\rm_{e}$ [\ion{O}{III}] is 557~K compared to the 3598~K for T$\rm_{e}$ [\ion{N}{II}], a consequence of the [\ion{N}{II}]~5755~\AA~line being fainter with higher uncertainty on its determined flux. While a number of PNe lie on the 1:1 line, a number of them have higher (T$\rm_{e}$ [\ion{N}{II}]) compared to their (T$\rm_{e}$ [\ion{O}{III}]) estimate. However, the PNe which do not lie on the 1:1 line (within error) have higher mean uncertainty on the T$\rm_{e}$ [\ion{N}{II}]  of 4589~K compared to those on the 1:1 line with a mean uncertainty of 3217~K, implying that only the more uncertain T$\rm_{e}$ [\ion{N}{II}] values do not lie on the 1:1 line. Furthermore, no difference in noted between the high- and low- extinction PNe in Figure~\ref{fig:temp}. 

A physical contribution towards higher T$\rm_{e}$ [\ion{N}{II}] values compared to T$\rm_{e}$ [\ion{O}{III}] could be the presence of high density clumps in PNe \citep{Morisset17}. At high densities, emission of the [\ion{N}{II}]~6548,~6583~AA~nebular lines can be suppressed due to collisional de-excitation (while emission of the [\ion{N}{II}]~5755~\AA~auroral line is unaffected), and consequently T$\rm_{e}$ [\ion{N}{II}] is overestimated \citep{fang18}. {Another reason for the overestimated T$\rm_{e}$ [\ion{N}{II}] values can be the recombination excitation of the [\ion{N}{II}]~5755~\AA~line \citep{Liu00} which can not be corrected for in distant extragalactic PNe (due to the intrinsically faint and hence unobserved recombination lines). However, such recombination contribution affects the T$\rm_{e}$ [\ion{O}{III}] by less than 1\% \citep{Liu00}.}

{Being unaffected by the presence of high-density clumps and recombination lines,} T$\rm_{e}$ [\ion{O}{III}] when measurable should be preferred for PN electron temperature estimation. Thus combined with its availability for the entire sample of PNe studied in this work, the lower uncertainty and inherent reliability , we utilise T$\rm_{e}$ [\ion{O}{III}] for computing chemical abundances in this work.

\subsection{ICF correction for oxygen and argon abundance determination}
\label{app:icf}

\begin{figure}
        \centering
        \includegraphics[width=\columnwidth,angle=0]{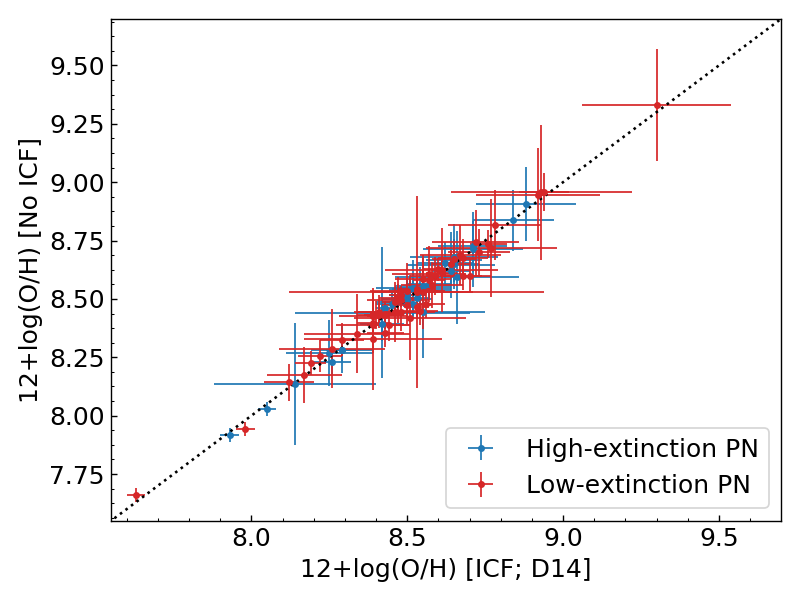}
        \caption{{For a sub-sample of PNe where \ion{He}{I} and \ion{He}{II} recombination lines are observed, we compare the oxygen abundance determined by NEAT following the ICF scheme of \citet{Delgado14} and compare with the oxygen abundance determined assuming no ICF correction. High- and low-extinction PNe are marked separately. The dotted line shows the 1:1 line.}}
        \label{fig:oxyicf}
\end{figure}

Ionic abundances of O$^{+}$ and O$^{++}$ are estimated from observed [\ion{O}{II}]~3726/3729~\AA~and  [\ion{O}{III}]~4363/4959/5007~\AA~line fluxes respectively. However, estimation of oxygen abundances may require correction for unobserved O$^{+3}$ ions. This can be achieved through the ICFs described by \citet{Delgado14} but that requires ionic abundance measurement of He$^{+}$ and He$^{++}$, which in turn requires observations of the relevant \ion{He}{I} and \ion{He}{II} recombination lines. {While such lines are observed for 86 of our 205 PNe, for the rest of our sample they remain unobserved. For these 86 PNe, Figure~\ref{fig:oxyicf} shows the oxygen abundance determined by NEAT following the ICF described by \citet{Delgado14} against the oxygen abundance assuming no ICF correction for these PNe. As seen from Figure~\ref{fig:oxyicf}, all points lie on the 1:1 line, within errors, thus the two abundance determination prescriptions lead to nearly identical oxygen abundance values. The ICF correction for unobserved O$^{+3}$ ions is thus negligible in our sample, as was also found by \citet{fang18} with deeper spectra of M~31 PNe. It is to be noted that the ICF correction for unobserved O$^{+3}$ ions can in theory be as high as $\sim10$ times of the observed ions but this case happens when the PN central stars have high effective temperatures, T$\rm_{eff}\sim200000$~K \citep[][see their Figure 5]{Delgado14}. Given that such high ICF corrections is not the case for the PNe in this work where \ion{He}{I} and \ion{He}{II} recombination lines were observed, it is also unlikely that such high temperature central stars are present in the PNe where such lines are not detected.} Thus, in this work we use the default NEAT prescription where the oxygen abundance is determined with ICF correction when He$^{+}$ and He$^{++}$ ionic abundances are measured and no ICF correction otherwise. We also note that ICF schemes from \citet{Delgado14} and \citet{KB94} give similar negligible corrections for oxygen abundance determination \citep{GarciaRojas16,fang18}.

\begin{figure}
        \centering
        \includegraphics[width=\columnwidth,angle=0]{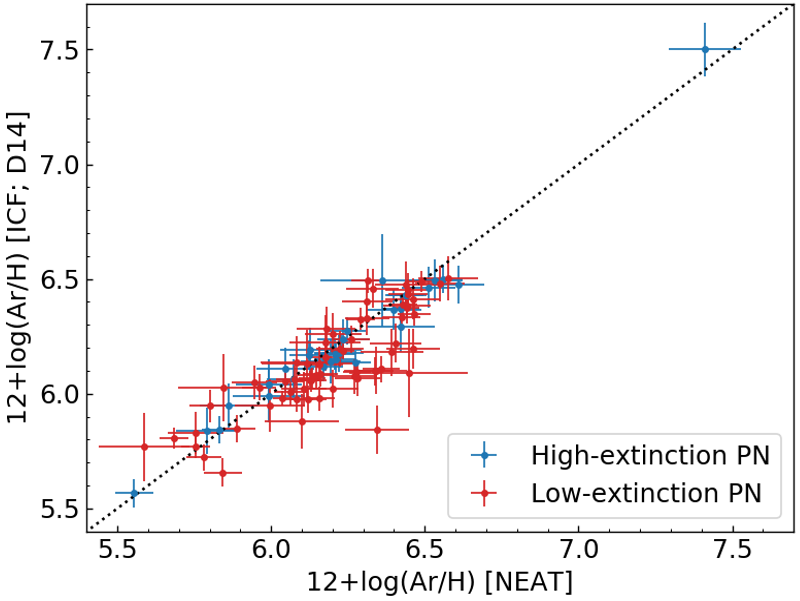}
        \caption{For a sub-sample of PNe where multiple argon ionic abundances have been determined, we compare the argon abundance determined by NEAT assuming no ICF correction and that determined by the ICF scheme of \citet{Delgado14} using only the Ar$^{++}$ ionic abundance. High- and low-extinction PNe are marked separately. The dotted line shows the 1:1 line.}
        \label{fig:icf}
\end{figure}

Ionic abundances of Ar$^{++}$, Ar$^{+3}$ and Ar$^{+4}$ are estimated from observed [\ion{Ar}{III}]~7136/7751~\AA, [\ion{Ar}{IV}]~4711/4740~\AA~and [\ion{Ar}{V}]~7005~\AA~line fluxes respectively. While the [\ion{Ar}{III}]~7136~\AA~line is observed for all 200 PNe where argon abundance is determined in this work, the other lines are observed for only a sub-sample of our PNe. The ICF scheme of \citet{Delgado14} is used for determining the argon abundance when only Ar$^{++}$ ionic abundance is measured. However, as this ICF scheme does not incorporate Ar$^{+3}$ and Ar$^{+4}$ ionic abundances, the default prescription from NEAT does not carry out an ICF correction for argon abundance determination when multiple ionic species have their abundances determined. In Figure~\ref{fig:icf}, we show that for the sub-sample of PNe where multiple argon ionic species are observed, the argon abundance determined by NEAT is consistent with that determined by applying the ICF scheme of \citet{Delgado14} to only the Ar$^{++}$ ionic abundance. This is true for both high- and low-extinction PNe. Thus, whether only the [\ion{Ar}{III}]~7136~\AA~line flux is measured or other forbidden argon lines fluxes are measured, we consistently determine the argon abundance in this work. We also note that ICF schemes from \citet{Delgado14} and \citet{KB94} give similar corrections for argon abundance determination \citep{GarciaRojas16,fang18}.

\subsection{Comparison of oxygen abundance estimates from this work with literature values}
\label{app:lit}

\begin{figure}
        \centering
        \includegraphics[width=\columnwidth,angle=0]{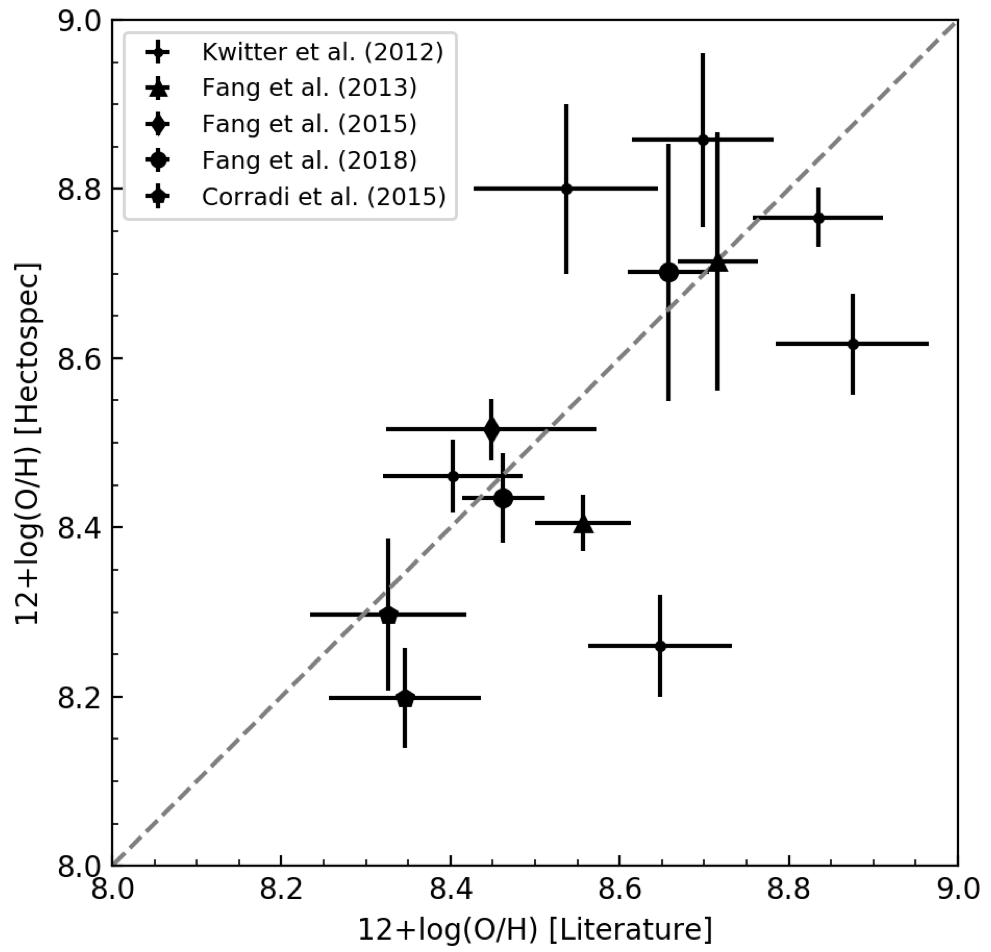}
        \caption{Comparison of the oxygen abundances {derived in this work} with those {computed} in the literature for the PNe in common. The PNe studied by different authors are marked with different symbols. The dotted line shows the 1:1 line.}
        \label{fig:comp_oh}
\end{figure}

Figure~\ref{fig:comp_oh} compares the oxygen abundances {derived} for PNe observed in this work which have oxygen abundances already published in the literature \citep{Kw12, corradi2015, fang13, fang15, fang18}. While most of the oxygen abundances values agree with each other, some scatter is observed. {The scatter is particularly for the oxygen abundances determined in PNe by \citet{Kw12} who employed 1D-CLOUDY \citep{cloudy} photoionisation models to determine chemical abundances without relying on line-ratios.} Note that the previously largest sample of PN abundances in M~31 ($\sim$50 PNe) were {computed} by \citet{san12} but since those have been reanalysed in this work, we do not show their previous {estimates} in Figure~\ref{fig:comp_oh}.

\section{AGB evolution and possible dependencies of PN oxygen abundances}
\label{sect:agb}

While the {derived} argon abundances in PNe have been found to be invariant during the AGB evolution, thus reflecting the ISM abundance at the time of their birth, AGB evolution effects have been suggested to modify the oxygen abundance {estimated} in the nebula from that of the progenitor star, for specific PNe \citep{Delgado-Inglada15,Garcia-Hernandez16}. These effects depend on the progenitor mass and metallicity according to AGB theoretical evolution models such as those described in \citet{Ventura17}.

For PNe evolving from stars with initial mass $\geq3\rm M_{\odot}$, hot-bottom burning (HBB) may result in an oxygen depletion of up to $\sim0.2$ dex, while for PNe evolving from stars with initial masses of $1-2\rm M_{\odot}$ and Z$<0.008$, third dredge-up (TDU) effects may result in an oxygen enrichment of up to $\sim0.3$ dex \citep[e.g.][]{Garcia-Hernandez16, Ventura17}. In a small sample of 20 MW PNe, \citet[][]{Delgado-Inglada15} found that oxygen is enriched in MW PNe with Carbon-rich (circumstellar) dust (CRDs), by up to $\sim0.3$ dex for intermediate metallicities of 12+(O/H) = 8.2--8.7, while oxygen is invariant in MW PNe with oxygen-rich (circumstellar) dust (ORDs). 

\begin{figure}
        \centering
        \includegraphics[width=\columnwidth,angle=0]{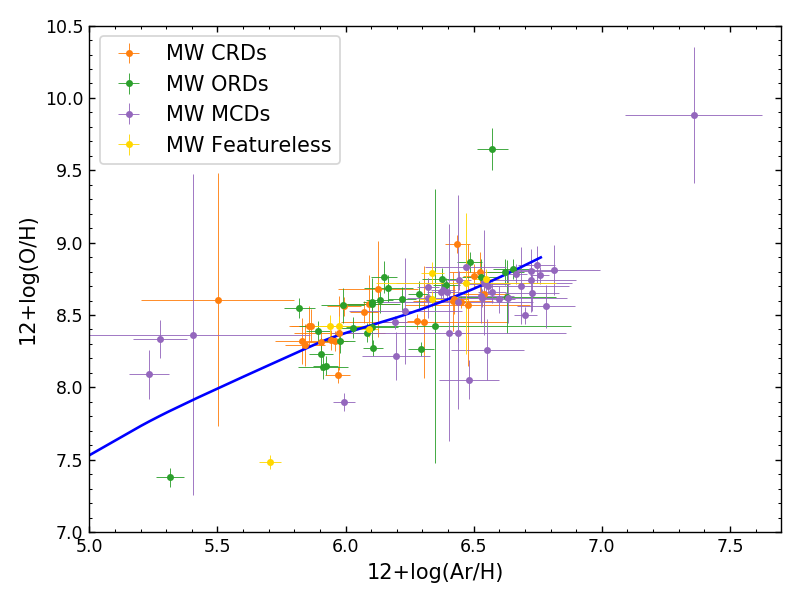}
        \caption{Oxygen vs. argon abundance distribution of 101 MW PNe marked by their circumstellar dust types from \citet{Ventura17}. The chemical evolution model track for the solar neighbourhood is from \citet{Kobayashi20}.}
        \label{fig:sample2_dust}
\end{figure}

We therefore check whether the oxygen abundance values of a larger sample of MW PNe depend on the PN circumstellar dust composition.  Figure~\ref{fig:sample2_dust} shows the distribution of oxygen abundances against the argon abundances of 101 MW PNe whose dust properties and abundances were tabulated by \citet[][their Sample 2]{Ventura17}. CRDs, ORDs as well as those PNe with featureless dust distribute around the ISM model predictions tracks for the solar neighbourhood (see \citealt{Kobayashi20} for details of the models). In particular, we compare the oxygen abundance of the PNe with that of the solar neighbourhood model at their given argon abundance. We find the MW CRDs and ORDs have mean offset in oxygen abundance of 0.06~dex ($\sigma\rm_{offset}=0.26$~dex) and 0.05~dex ($\sigma\rm_{offset}=0.41$~dex) respectively against the model. The featureless dust MW PNe have a mean offset of -0.02~dex ($\sigma\rm_{offset}=0.43$~dex) against the model. The offsets are much lower than the mean uncertainty, which is $\sim$0.19~dex. The errors on the {computed} oxygen and argon abundances of the M~31 disc PNe in the current work are comparable to those of the MW PNe sample compiled by \citet{Ventura17}, as well as by \citet{Delgado-Inglada15} for their statistically limited sample. We conclude that there is no segregation of CRD/ORD (as well as featureless) MW PNe in Figure~\ref{fig:sample2_dust}.

We note also in the same figure that many of the MW PNe with mixed chemistry dust (MCDs) that are metal-rich (12 + log(Ar/H) > 6.3) preferentially have log(O/Ar) values below the model tracks, indicating lower oxygen or oxygen depletion. These MCDs have a mean offset of -0.13~dex ($\sigma\rm_{offset}=0.21$~dex) against the model. These PNe have been suggested by \citet{GarciaHernandez14} to be the  youngest (<300 Myr), most metal-rich PNe in the MW sample. These PNe likely evolve from the most massive progenitors that display HBB, as predicted by the AGB evolution models discussed in \citet{Ventura17}. Thus, the log(O/Ar) values of MW PNe younger than $\sim$300 Myr may be reduced due to oxygen depletion (see Figure~\ref{fig:sample2_dust}). The M~31 low- and high- extinction PNe have average ages $\sim4.5$~Gyr and $\sim2.5$~Gyr respectively with the bulk of the latter having likely formed in a burst of star formation $\sim$2 Gyr ago (\citetalias{Bh+19b}). This implies that a very small number of PNe with very young massive progenitors (affected by HBB) are expected in our sample.  

To summarise, we find no conclusive evidence of AGB evolution effects with modification of the oxygen abundance in the M~31 disc PNe studied in this work. Any such effect is within the measurement errors. We thus conclude the oxygen abundances derived for M~31 PNe reflect their birth ISM chemical abundances, within the errors.


\bsp	
\label{lastpage}
\end{document}